\documentclass[useAMS,usegraphicx,usenatbib,a4paper]{mn2e}
\usepackage[T1]{fontenc}
\usepackage{aecompl}
\usepackage[dvips, bookmarks, pdfborder={0 0 0.1}]{hyperref}
\usepackage[]{amsmath}
\usepackage{mathrsfs}
\usepackage{aas_macros}
\usepackage{color}
\usepackage{mn2e-bst} 

\newcommand{\VR}{\vec{r}}
\newcommand{\VV}{\vec{v}}
\newcommand{\D}{\mathrm{d}}

\graphicspath{{figures/}}

\newcommand{\myplot}[1]{\includegraphics[width=0.5\textwidth]{#1}}
\newcommand{\myplottwo}[2]{\myplot{#1}\myplot{#2}}

\newcommand{\mytab}{\begin{table}[htb]}
\newcommand{\myfig}{\begin{figure}[htbp]}

\newcommand{\mybibstyle}{mymn}

\begin{document}
\title[Dynamical inference with steady-state tracers]{The orbital PDF: general inference of the gravitational potential from steady-state tracers}
\author[J. Han et al.]{Jiaxin Han,$^{1}$\thanks{jiaxin.han@durham.ac.uk} 
Wenting Wang,$^1$ Shaun Cole,$^1$ Carlos S. Frenk$^1$\\
$^1$Institute for Computational Cosmology, Department of Physics,
Durham University, South Road, Durham, DH1 3LE, UK\\
}
\maketitle

\begin{abstract}
 We develop two general methods to infer the gravitational potential of a system using steady-state tracers, i.e., tracers with a time-independent phase-space distribution. Combined with the phase-space continuity equation, the time independence implies a universal Orbital Probability Density Function (oPDF) $\D P(\lambda|{\rm orbit})\propto \D t$, where $\lambda$ is the coordinate of the particle along the orbit. The oPDF is equivalent to Jeans theorem, and is the key physical ingredient behind most dynamical modelling of steady-state tracers. 
 In the case of a spherical potential, we develop a likelihood estimator that fits analytical potentials to the system, and a non-parametric method (``phase-mark'') that reconstructs the potential profile, both assuming only the oPDF. The methods involve no extra assumptions about the tracer distribution function and can be applied to tracers with any arbitrary distribution of orbits, with possible extension to non-spherical potentials. The methods are tested on Monte Carlo samples of steady-state tracers in dark matter haloes to show that they are unbiased as well as efficient. A fully documented \textsc{C/Python} code implementing our method is  freely available at a GitHub repository linked from  \url{http://icc.dur.ac.uk/data/#oPDF}.
\end{abstract}


\section{Introduction}
Since dark matter does not emit or absorb electromagnetic radiation, gravitational modelling is of fundamental importance to the determination of the dark matter distribution. Such modelling can be performed using either gravitational lensing \citep[e.g.,][]{Bartlemann10,GAMA}, or the dynamics of tracers (e.g., stars or galaxies; see \citealp{MassRev} for a recent review on galaxy mass inferences). 

One straightforward way to perform dynamical modelling is to fit a
proposed phase-space distribution function (DF) to the observed
positions and velocities of tracer particles. Thanks to Jeans theorem
\citep[see e.g.][]{BT08}, which states that functions of the integrals
of motion, $J$,  are solutions of the Boltzmann equation, one can
simply consider DFs of the form $f(J)$. Under certain conditions, one can invert the observed density profile, $\rho=\int f(J) \D^3v $, of
the tracer to construct a specific family of $f(J)$ (e.g., \citealp{Eddington,Osipkov79,Merritt85,Cudderford}; hereafter referred to as density profile inversion). The density profile inversion depends on the potential of the halo, and the resulting $f$ is thus potential dependent. However, some further assumptions about the functional form of $f(J)$ are required to perform the inversion. These
assumptions are typically motivated either empirically
\citep[e.g.,][]{Wojtak,WE15b},  or by mathematical simplicity
\citep[e.g.,][]{Evans06}. When proposing the DFs, one is free to choose the integrals of motion, to be either classical integrals such as energy and angular momentum, or the more theoretically appealing actions~\citep[see e.g.,][for such recent models on halo stars]{Posti15,WE15}.

In \citet{Wilkinson99} and \citet{Wang},
solutions of the form $f(E,L)=f(E)L^{-2\beta}$ were used to constrain
the potential of the Galactic halo, where $E$ and $L$ are the energy
and angular momentum of tracer particles. In particular, \citet{Wang}
applied the $f(E,L)$ method to mock stellar haloes~\citep{Andrew} constructed from
the Aquarius simulations of $\Lambda$CDM galactic haloes~\citep{Aquarius} and found significant biases
in the fitted masses. These biases suggest that the proposed $f(E,L)$ DF
does not describe well the observed phase-space distribution of the
mock stars. However, it is not clear whether the discrepancy is due to
departures from dynamical equilibrium by the stars within the halo potential, or to the
lack of generality in the proposed $f(E,L)$ functional form used to
describe the distribution~\citep[e.g.,][]{BM82}. In the former case, the stars would represent an intrinsically biased tracer, and the modelling of their distribution would not give the correct potential. In the latter case, one can still hope to find a DF that fits the observed distribution with the correct potential once the extra assumptions in the functional form of $f(E,L)$ are relaxed or removed. 

A method requiring no assumptions on the form of $f(E,L)$ is achievable, which is what we develop in this work. The starting point of our method is the definition of tracer. We define a tracer as a population of objects whose phase-space DF does not evolve with time (i.e., is in a steady state), so that modelling their DF at an arbitrary time is generally possible and useful. It immediately follows from this definition that the probability of observing a particle at a position on its orbit is proportional to the time it spends near that position, i.e., $\D P|{\rm orbit}\propto \D t$. Formally, this can be shown to be a result of phase-space continuity (Section~\ref{sec:equiv_continuity}). We give a thorough description on this Orbital Probability Density Function (oPDF) in Section~\ref{sec:Tracer}, with further discussion in Section~\ref{sec:discussion_tracer}. 

This simple relation actually contains all the information required to
model the potential of the system. We demonstrate this by constructing
explicit estimators for the potential from the oPDF in
Section~\ref{sec:estimator}. Expressed in action-angle coordinates,
where the angles evolve uniformly over time, the steady-state
distribution is a uniform distribution in angle, also known as
the orbital roulette~\citep[][hereafter BL04]{Roulette}. Two
minimum-distance estimators have been proposed in BL04 to infer the
potential using the uniform angle distribution. Unfortunately, when
applied to a $\Lambda$CDM halo potential, we find that
these phase angle estimators only probe the gravity at (or equivalently, the halo mass inside) a tracer-specific
characteristic radius of the Navarro-Frenk-White~\citep[NFW;][]{NFW96,NFW97} type potential, resulting in a high
degeneracy in the mass ($M$) and concentration ($c$) parameters of the
halo potential. 

Expressed in the radial coordinate, the steady-state
distribution translates into an orbit-dependent DF
in $r$. For any given potential, one can then predict a radial
distribution for the tracer according to the occurrence of orbits in
the data. A likelihood estimator can be constructed by comparing the
observed radial distribution to the predicted distribution. This
radial likelihood estimator is largely able to break the degeneracy in
$M-c$ and provides a good constraint on the halo potential profile over
a much larger radial range. This parametric likelihood estimator is described in Section~\ref{sec:like}.

Alternatively, the degeneracy in the phase angle estimators can be utilized to break the degeneracy itself. In particular, the degeneracy in the mean-phase estimator is so strong that it provides no constraint on the halo mass profile anywhere except at the characteristic radius, leaving the shape of the halo mass profile unconstrained. Such a degeneracy can be broken by applying the estimator multiple times to subsamples of the tracer in different radial ranges, thus constraining simultaneously the halo mass at different characteristic radii. At the same time, the perfect degeneracy means that profiles of any shape can be adopted to fit for the characteristic mass. Fitting two profiles of different shapes with the mean-phase estimator, we can find out the characteristic mass point by locating the point where the two mass profiles intersect. This leads to our non-parametric potential profile reconstruction method, in which we fit two elementary one-parameter profiles with the mean-phase estimator to ``mark out'' the characteristic mass point in each radial bin. This ``phase-mark'' method is detailed in section~\ref{sec:mark}.


The oPDF describes the conditional distribution of a particle in phase
space given its orbit. Coupled with assumptions on the prior
distribution of orbits, one can recover a full phase-space
DF applying Bayes' theorem. In
Section~\ref{sec:DFconnection}, we will show that these distribution
functions are fully compatible with those constructed from a density profile
inversion. Such a DF can then be used to fit the
observed distribution of the tracer to infer the potential, which is the approach
taken by conventional DF methods. If the assumptions on the distribution of orbits are correct, then the DF method is fully
compatible with ours. However, our method still works even if these
assumptions fail, while the validity of the DF methods is
intrinsically limited by the validity of these assumptions. 

There are some DF methods based on more general
assumptions about the distribution of orbits. For example,
\citet{Bovy} generalized the Roulette distribution to
a Bayesian likelihood estimator by combining the uniform distribution
of action-angles with the distribution of orbital parameters (e.g.,
$(E,L)$). The latter is modelled parametrically or with histograms, and
the parameters of the distribution of orbits are further marginalized
over some assumed priors. They applied their method to infer the
potential of the Solar system using the planets as
tracers. \citet{Magorrian} also proposed a Bayesian method by
modelling the distribution of orbits non-parametrically with an
arbitrary number of Gaussians in action space, and then marginalizing
over the proposed prior distribution of the normalization, location
and width of the Gaussians. These methods are still not
assumption-free, because a particular form for the distribution of
orbits and priors for their parameters still need to be assumed. 
Adopting more general functions to describe the distribution
of orbits also tends to complicate the mathematical and computational
aspects of the problem tremendously. Compared with these Bayesian
marginalization methods, our method is much simpler and more
intuitive. We do not need to model the distribution of orbits
at all, so our method is truly assumption-free in so far as the distribution of orbits is concerned.

Our likelihood estimator is closely related to Schwarzschild's
method \citep{Schwarzschild}, a general numerical method that solves
the $\rho=\int f(J) \D^3v$ equation numerically to obtain $f$ as well
as the potential. Without loss of generality, our method
effectively works by determining one orbit from the phase-space
coordinate of each particle, avoiding the numerical search for the
combinations of orbits used in Schwarzschild's method. 
We elaborate on this point in Section~\ref{sec:Schwarzschild}.

In a follow up paper~\citep[][Paper~II]{paperII}, we apply our oPDF analysis to
tracers of Galaxy-sized haloes constructed from the Aquarius simulations~\citep{Aquarius}, to study the
dynamical status of both the dark matter and stars in the halo, and to
gain insights on the intrinsic uncertainties in the inferred dynamical
mass of the Milky Way.

\section{Steady State Tracers}\label{sec:Tracer}
As the fundamental concept used in this work, we start by deriving the
steady-state distribution of test particles, which is used as the
definition of tracers throughout.

\subsection{The orbital Probability Density Function}
We consider steady-state systems consisting of a set of test particles
(a tracer) moving under a gravitational potential. We require both the
total potential and the phase-space distribution of the tracer particles to be
in a steady state, i.e., to not evolve with time. As long as the total
potential is static, we do not care whether it is generated by the
tracer alone or purely by an external field, with the tracer being
massless, or from both components. The static potential assumption is
reasonable as long as the crossing time for tracer particles is much
smaller than the timescale for the variation in potential.

Under the static potential condition, each particle has a fixed and
predictable orbit. If the tracer particles are to be in a steady state,
then for any given orbit, the probability of observing one particle at
a given position (labelled by some parameter $\lambda$) has to be proportional to the time it spends at
that position, i.e.,
\begin{equation}\label{eq:oPDF}
 \D P(\lambda|{\rm orbit})/\D\lambda \propto \D t(\lambda|{\rm orbit})/\D\lambda.
\end{equation} In other words, if each particle has a fixed orbit, then the travel time on different parts of the orbit determines the density of particles observed along the orbit. We call equation~\eqref{eq:oPDF} the orbital Probability Density Function (oPDF). This can be understood as arising from the ergodicity of each particle in the steady-state system. The static potential leads to fixed orbits. If the overall system is in a steady state, ergodicity translates each orbit into a PDF.

\subsection{oPDF from phase-space continuity}\label{sec:equiv_continuity}
Formally, we can derive the oPDF from the phase-space continuity
equation (i.e, the collisionless Boltzmann equation) of the steady-state system as shown below. Readers not interested in the proof can skip
this subsection.

Let us consider the two requirements of our steady state system:
\begin{description}
 \item[\textit{Static Potential.}] Since the potential is fixed, each
 particle has a fixed and predictable orbit. Formally, one would be
 able to specify the phase-space coordinates with a set of orbital
 parameters ${Q_i} (i=1...n)$ determining the shape of the orbit, plus
 one affine parameter, $\lambda$, specifying the current position of the particle
 along the orbit. Note that $\lambda$ can be any parameter that uniquely specifies the position of the particle on the given orbit, e.g., radius $r$, velocity $v$ or the elapsed time since apocentre. Then the phase space of tracer particles is fully
 sampled by the distribution of orbits and the current position of the
 particles on each orbit. The distribution of orbits is fixed in a
 collisionless system, and any evolution of phase-space density is
 only caused by the change of the on-orbit position of each
 particle. In this coordinate system, the phase-space continuity
 equation reads
 \begin{equation}
  \frac{\partial f}{\partial t}+\sum_i \frac{\partial(f\dot{Q_i})}{\partial Q_i} + \frac{\partial (f\dot{\lambda})}{\partial \lambda}=0.
 \end{equation}
 Since $\dot{Q_i}=0$, we have
 \begin{equation}\label{eq_continuity}
  \frac{\partial f}{\partial t}+\frac{\partial (f\dot{\lambda})}{\partial \lambda}=0.
 \end{equation}

 \item[\textit{Steady-state.}] To be able to predict the distribution
 of particles we require a tracer population to be in a steady state,
 that is, $\partial f/\partial t=0$ at any point in phase
 space. Immediately, from equation~\eqref{eq_continuity} this implies
 $\partial{(f\dot{\lambda})}/{\partial \lambda}=0$ and hence
 \begin{equation}
  \frac{\D P}{\D \lambda}|{Q} \propto f({Q},\lambda)\propto \frac{1}{\dot{\lambda}},
 \end{equation}
  where ${Q}$ denotes the set of orbital parameters. That is, the
  probability of observing a particle at a given position is
  proportional to the time it spends near that position:
 \begin{equation}
  \D P|{Q}\propto \frac{\D\lambda}{\dot{\lambda}}=\D t.
 \end{equation}
\end{description} Q.E.D.

Note this is phase-space continuity and is more general than configuration-space continuity for steady flows.
The oPDF is a fundamental equation governing the distribution of
steady-state tracers. It is a very general result that follows from very general
assumptions.  We will also call this phase-space steady-state
distribution the equilibrium distribution. Note that the definition of
a tracer puts no constraint on the distribution of orbits. A tracer
with any arbitrary distribution of orbital parameters can be
constructed, as along as the oPDF is satisfied.

\subsection{oPDF in spherical systems}
In this work we focus the application of the oPDF to a spherically symmetric
potential, whose value depends only on radius
$\psi(r,\theta,\phi)=\psi(r)$. In this conservative central force
field, the binding energy,
\begin{equation}
 E=-(\frac{v_r^2}{2}+\frac{v_t^2}{2}+\psi(r)),
\end{equation}
and angular momentum, 
\begin{align}
\vec{L}&=\vec{r}\times \vec{v}\\
&=r v_t\vec{e}_L,
\end{align}
of each particle are conserved, and form a complete set of orbital
parameters. 
Taking $r$ as the affine parameter $\lambda$ along the orbit,
equation~\eqref{eq:oPDF} becomes
\begin{equation}\label{eq:oPDFr}
 \D P(r|E,L)=\frac{\D t}{\int \D t}=\frac{1}{T}\frac{\D r}{|v_r|},
\end{equation}
where $T=\int \D t$ is the period of the orbit. Note that we only need
$L$ rather than $\vec{L}$ if we are only interested in the radial
motion of particles. Since the orbit is symmetric for the inward and
outward-going parts, we ignore the direction of the radial velocity
and take one single journey between pericentre $r_{\rm p}$ and apocentre
$r_{\rm a}$ as one period. When radial cuts are
imposed, we only need to replace the orbital limits $r_{\rm p}$ with
$\max(r_{\rm p}, r_{\rm min})$ and $r_{\rm a}$ with $\min(r_{\rm a},
r_{\rm max})$, since
equation~\ref{eq:oPDF} holds within any radial range.

More generally, in equation~\eqref{eq:oPDF} the choice of the position
variable, $\lambda$, is not limited to the radial coordinate. It can be any
variable that uniquely determines the phase-space coordinate on its
orbit. As an example, we can choose $\lambda$ to be the travel time a
particle has spent to get to the current position. In this coordinate,
the distribution of particles is uniform along the orbit. If we define an
angle at each position $r$ as
\begin{equation}\label{eq_theta}
\theta(r)=\frac{\int_{r_{\rm p}}^r dr/|v_r|}{T},
\end{equation} where $r_{\rm p}$ is the pericentre distance, then the orbital PDF becomes
\begin{equation}
 \D P(\theta|E,L)=\D\theta.
\end{equation} 
This PDF is a uniform distribution, with $\theta \in [0,1]$. This
angle is known as an action-angle, and its randomness has been argued for or
assumed in previous works \citep[``random phase principle'' in
BL04;][]{Bovy}. Here we do not assume randomness of the angle; 
rather the randomness is a derived property from the continuity
equation of the steady-state system coupled to the uniform time
evolution of the action-angle. We will call this angle the radial
phase.

Despite the focus on spherically symmetric potentials in this work,
the oPDF does \emph{not} require spherical symmetry for the \emph{tracer}
distribution. For example, the oPDF holds for a tracer on a single
elliptical orbit with the same $(E,L)$. The method can be generalized
to $(E,\vec{L})$ orbits where the asphericity of the orbits is
explicitly used.

\subsection{Equivalence to Jeans theorem and connection to other DFs}\label{sec:equiv_Jeans}
A fundamental constraint on the DF of steady-state systems is provided by the Jeans
theorem. It is useful to clarify how it connects to the oPDF.

Below we demonstrate the connection in a spherically symmetric
system. In such a system, ignoring the angular distribution, each
particle has three phase-space coordinates, which can be specified by
$(r,v_r,v_t)$ or equivalently $(E,L,r)$. The phase-space distribution
function can generally be written as
\begin{align}
\D P(\vec{r},\vec{v})&=f(E,L,r) \, \D^3 r \, \D^3 v\\
&=f(E,L,r) \, 8\pi^2 L \, \frac{\D r \, \D E \, \D L}{|v_r|}. 
\label{eq_PRaw} 
\end{align}
Now we prove the equivalence of oPDF with Jeans theorem. If Jeans theorem holds, i.e, $f(E,L,r)=f(E,L)$, then
\begin{align}
 \D P(r|E,L)&=\frac{\D P(E,L,r)}{\int_r \D P(E,L,r)}\\
 & \propto \frac{\D r}{|v_r(E,L,r)|}.
\end{align}
Conversely, if the oPDF holds, then
\begin{align}
 \D P(E,L,r)&= \D P(E,L)\, \D P(r|E,L)\\
 &= \frac{\D^2 P(E,L)}{\D E \D L} \frac{\D r}{|v_r|T} \, \D E \, \D L .
\end{align}
Combining with Eq.~\eqref{eq_PRaw}, we have 
\begin{equation}
f(E,L,r)= \frac{1}{8\pi^2 L T(E,L)}\frac{\D^2 P(E,L)}{\D E \D L},
\end{equation}
 which is purely a function of $(E,L)$. Q.E.D. 

Put simply, the Jeans theorem implies a known radial distribution, so
that the full phase-space DF only needs to be
specified in $(E,L)$ coordinates.

Starting from the oPDF, one can construct the full DF from Bayes' theorem, by specifying the distribution of orbits, $P(E,L)$, of the tracer, as
\begin{equation}\label{eq:Pfull}
\D P(\VR,\VV)=\D P(r|E,L) \, \D P(E,L) .
\end{equation}
Depending on how $P(E,L)$ is specified, the constructed DF varies. A
popular way of specifying $P(E,L)$ relies on the radial profile
constraint
\begin{equation}\label{eq:RadialConstraint}
 \int \frac{\D P}{\D^3r \D^3v}\, d^3v=\rho(r) .
\end{equation}
This is obtained mathematically through, for example, an Abel transform, with
$\rho(r)$ being the parametrized density profile of the tracer. Even
though Eq.~\eqref{eq:Pfull} is not explicitly used, its equivalent,
Jeans theorem is used to propose a DF of the form $f(E,L)$. However,
at this stage, the mathematical inversion of
Eq.~\eqref{eq:RadialConstraint} is not generally solvable without further restrictions, and one
typically needs to further assume some more specific forms of $f(E,L)$,
for example $f(E,L)=L^{-2\beta}F(E)$ \citep{Camm,Cudderford,Wilkinson99,Wang}.  In some
other works, the radial constraint is not used and a more general
distribution of orbits is proposed \citep{Bovy, Magorrian}.

Note the distribution function constructed following Eq.~\eqref{eq:Pfull} is non-negative as long as $\D P(E,L)$ is non-negative, because $\D P(r|E,L) \propto dt \ge 0$ always holds. As we will see later, in practice we approximate the $\D P(E,L)$ with the empirical distribution given by Eq.~\eqref{eq_ELdiscrete}, which is always non-negative and corresponds to the discrete realization of a physical DF.

\section{Data: ideal tracers}\label{sec:data}
In order to test the performance of potential estimators, 
we first generate a set of Monte-Carlo steady-state
tracers. Tracer particles are generated according to the probability
distribution
$\D P(\VR,\VV)=f(E,L)\,\D^3r\, \D^3v$ used in
\citet{Wang}. This DF is constructed by inverting a
double-power law tracer density profile, assuming
$f(E,L)=f(E)L^{-2\beta}$ and a Navarro-Frenk-White~\citep[NFW][]{NFW96,NFW97} potential. It describes a
spherically symmetric steady-state system of tracers inside an {\sc nfw}
halo. The detailed form of the DF is complicated
and we refer the reader to Eq.~12 of \citet{Wang} for a full
description.
The parameters of the DF include the mass, $M$, and
concentration, $c$, of the {\sc nfw} halo, the tracer velocity
anisotropy, $\beta$, and the double power law slopes and pivot radius of the tracer
density profile, $\alpha$, $\gamma$ and $r_{\rm p}$. Their values are chosen
to best match the distribution of mock stars inside a Milky Way (MW) 
sized halo in the Aquarius simulation~\citep{Andrew}, with $M=1.83\times10^{12}\,\mathrm{M}_\odot, c=16.2,\beta=0.715,r_{\rm p}=69.0\,{\rm kpc},\alpha=2.30,\gamma=7.47$. The mock tracers generated according to this
$f(E,L)$ are expected to be a self-consistent, yet simplified,
realization of the distribution of stars in the MW. Tracer particles are generated between $1$ and $1000\,{\rm kpc}$ in radius. We will
call these catalogues ideal tracers. Since it is a steady state
system, the oPDF is applicable.\footnote{Strictly speaking, the DF of \citet{Wang} only describes a system of massless tracers in an external NFW potential, because the tracer density would exceed the total density as $r\rightarrow 0$ unless the tracers were massless. However, as we state at the beginning of section~\ref{sec:Tracer}, our definition of a steady-state system does not depend on the origin of the potential and equally applies to massless tracers in an external potential. So this is not an issue for our analysis. It may also be worth noting that the tracer velocity dispersion approaches $0$ at $r=0$ for this DF according to \citet{AE09}. However, this does not prevent the system  being in a steady-state. In addition, our radial cut of $1$ to $1000\,{\rm kpc}$ ensures we are not affected by the behaviour at the centre.} In fact, as we have discussed in
Section~\ref{sec:equiv_Jeans}, any $f(E,L)$ DF has
to be compatible with the oPDF but it also imposes additional assumptions.

In real observations, the phase space coordinates of tracer particles are inevitably affected by observational errors. We do not consider such errors in the main portion of this paper. In the following we simply use the ideal tracer with their exact phase space coordinates, to test our methods. We briefly discuss possible extensions of applying our methods to real data in Section~\ref{sec:obs}.

\section{Parametric Potential Estimators}\label{sec:estimator}
The problem we are trying to solve is: given a tracer population in equilibrium, with
observed positions and velocities, how do we infer the potential of the
steady-state system in which it resides? For any assumed potential, we can convert the
positions and velocities of particles into $(r,E,L)$ or $\theta$
coordinates. This results in empirical distributions for the
particles in these coordinates. By comparing these distributions with
the oPDF expected for tracer populations it is possible to 
constrain parameters of the underlying potential. Below we consider 
various parameter estimators that compare  empirical and
expected distributions: two minimum distance estimators given in BL04,
and one maximum likelihood estimator that we develop in this work.

\subsection{Minimum Distance Estimators and Parameter Degeneracies}
For minimum distance estimators, one constructs a metric to specify
 the distance between the empirical and theoretical distributions, and
 minimizes this distance to infer model parameters. Since the phase
 angles are computed quantities assuming a model potential rather than observed quantities, one cannot construct likelihood estimators
 using the distribution of the angles (the differentiation of angles
 would introduce model dependence). Following BL04, we consider two
 distance measures to quantify the consistency of the data with a uniform
 phase angle distribution. For a sample of $N$ particles drawn from  a
 uniform phase distribution, the mean phase, $\bar{\theta}$, is
 expected to follow a normal distribution with mean $0.5$ and standard
 deviation $1/\sqrt{12N}$ according to the central limit theorem. The
 normalized mean phase deviation,
\begin{equation}
 \bar{\Theta}=\sqrt{12N}(\bar{\theta}-0.5), \label{eq:meanphase}
\end{equation}
is then a standard normal variable. $\bar{\Theta}^2$ then serves as a
measure of the distance of the actual phase distribution from the
expected uniform distribution. If the data follows the model
distribution, then the $\bar{\Theta}^2$ from different realizations of
the same distribution will follow a $\chi^2$ distribution with one degree of freedom. Hence, the
discrepancy level of the data from the model can be quantified by the
probability of obtaining a $\chi^2$ as extreme as the measured value of
$\bar{\Theta}^2$.

A more sophisticated distance measure can be constructed by comparing the cumulative data distribution, $P_{<\theta}$, to the expected distribution, $\hat{P}_{<\theta}=\theta$, as~(BL04)
\begin{equation}\label{eq:AD}
 D=\int_0^1\frac{(P_{<\theta}-\hat{P}_{<\theta})^2}{\mathrm{var}\left(P_{<\theta}\right)} \D\theta. 
\end{equation}
where
$\mathrm{var}\left(P_{<\theta}\right)={\theta(1-\theta)}/{N}$ is
the variance of $P_{<\theta}$. This is known as the Anderson-Darling
~\citep[AD][]{AD} distance measure. For a set of $N$ particles with phase angles, $\theta_i$, Eq.~\eqref{eq:AD} can be evaluated:
\begin{equation}
 D=-N+\frac{1}{N}\sum\limits_{i=1}^N (1-2i)\ln \theta_i -[1+ 2(N-i)] \ln (1-\theta_i).
\end{equation}
The above form is simpler than the one derived in BL04. The
theoretical distribution of $D$ can be found from Monte-Carlo
simulations. A fitting formula is given by BL04 which fits the tail of
the distribution. In Appendix~\ref{app:ADdist}, we provide a
binormal fit to the distribution of $\ln(D)$ that works very well for the full
DF.

A small value of $\bar{\Theta}^2$ or $D$ suggests a good match between
data and model. Hence, these two distance measures can be used to fit
the data to parametric models, by searching for parameters that
minimize the distances. Confidence intervals can also be defined by
distance contours chosen so that the probability of obtaining a
distance measurement as extreme as that of the contour value 
equals the desired confidence level.

\begin{figure}
 \myplot{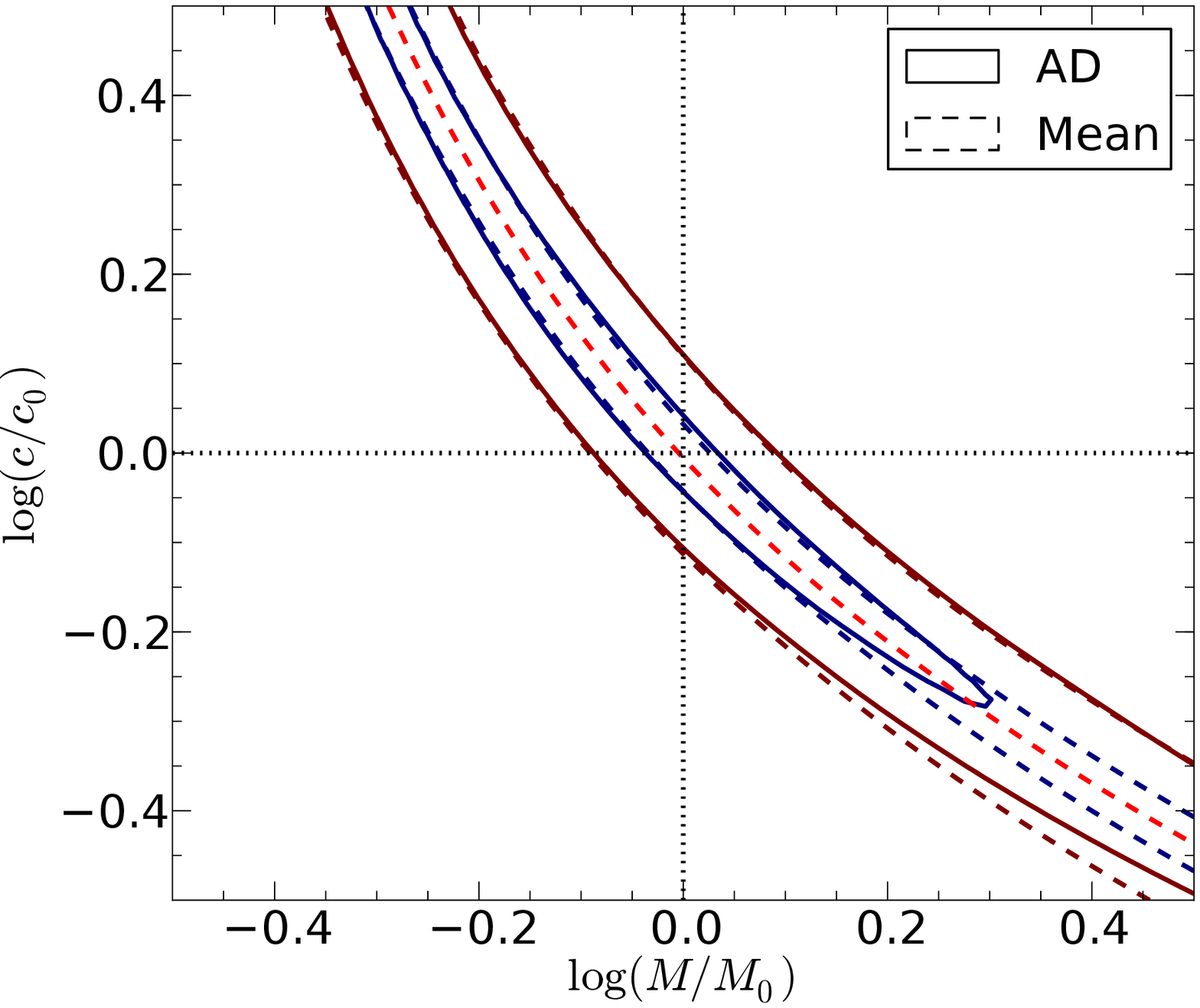}
 \caption{Constraints on the mass, $M$, and concentration, $c$,
   derived from 1000 particle realizations drawn from a model {\sc
   nfw} halo.  The outer (blue) and inner (brown) contours mark the 1 and 3$\sigma$
   confidence regions respectively, for the mean phase (dashed) and AD (solid)
   estimators. The central red dashed line marks $\bar{\Theta}=0$. The
   parameters are expressed in units of the true parameter values,
   $M_0$ and $c_0$.}\label{fig:MinDist}
\end{figure}

In Fig.~\ref{fig:MinDist} we apply the minimum distance estimators to
an ideal tracer of 1000 particles. It is obvious in
Fig.~\ref{fig:MinDist} that the mass-concentration parameters are
highly degenerate for both estimators. For the mean-phase estimator,
there is not a unique minimum distance point but a minimum distance
line with $\bar{\Theta}=0$, as marked by the central red dashed
line. As a result, there is not a single best-fitting parameter, but a
line of degenerate solutions. The AD estimator shows a slightly weaker
degeneracy, which can be understood because it uses more information
than the mean-phase estimator. However, the best-fitting parameters
from the AD estimator still depend sensitively on the initial
parameters due to the degeneracy. 


Even though the usefulness of these two statistics as estimators is
severely limited by the strong degeneracies, they can still be used as
theoretical probes to identify regions of discrepancy in heterogeneous
data. In particular, the signed mean phase deviation as a standard
normal variable is easy to calculate as well as being easy to interpret,
making it a good residual measure of the phase distribution under the
proposed potential. We demonstrate such an application in
Section~\ref{sec:marginalization}. Similar applications can be found
extensively in paper~II when analysing tracers from cosmological simulations. 


\subsection{Breaking the degeneracy: the radial likelihood estimator}\label{sec:like}
In addition to the minimum distance estimators in $\theta$-space, we also
try to construct maximum likelihood estimators. Since $\theta$ is
not a direct observable but is model dependent, its PDF cannot be
directly used to construct a likelihood. Instead, we work with the
directly observable $r$-coordinate to calculate the likelihood of
the observations, making use of the oPDF, $P(r|E,L)$. Since $E$ is unknown
before the potential is known, trying to use the conditional
probability ${\cal L}=\prod_i \D P(r_i|E_i,L_i)$ as a likelihood to infer the
potential will fail. Since $P(E,L)$ is solely a characteristic
property of the tracer (tracers with any arbitrary $P(E,L)$ can be
constructed or selected) and is independent of the potential, one could
seek to eliminate the $(E,L)$ dependence by marginalizing over their
prior distributions.

\begin{figure}
 \myplot{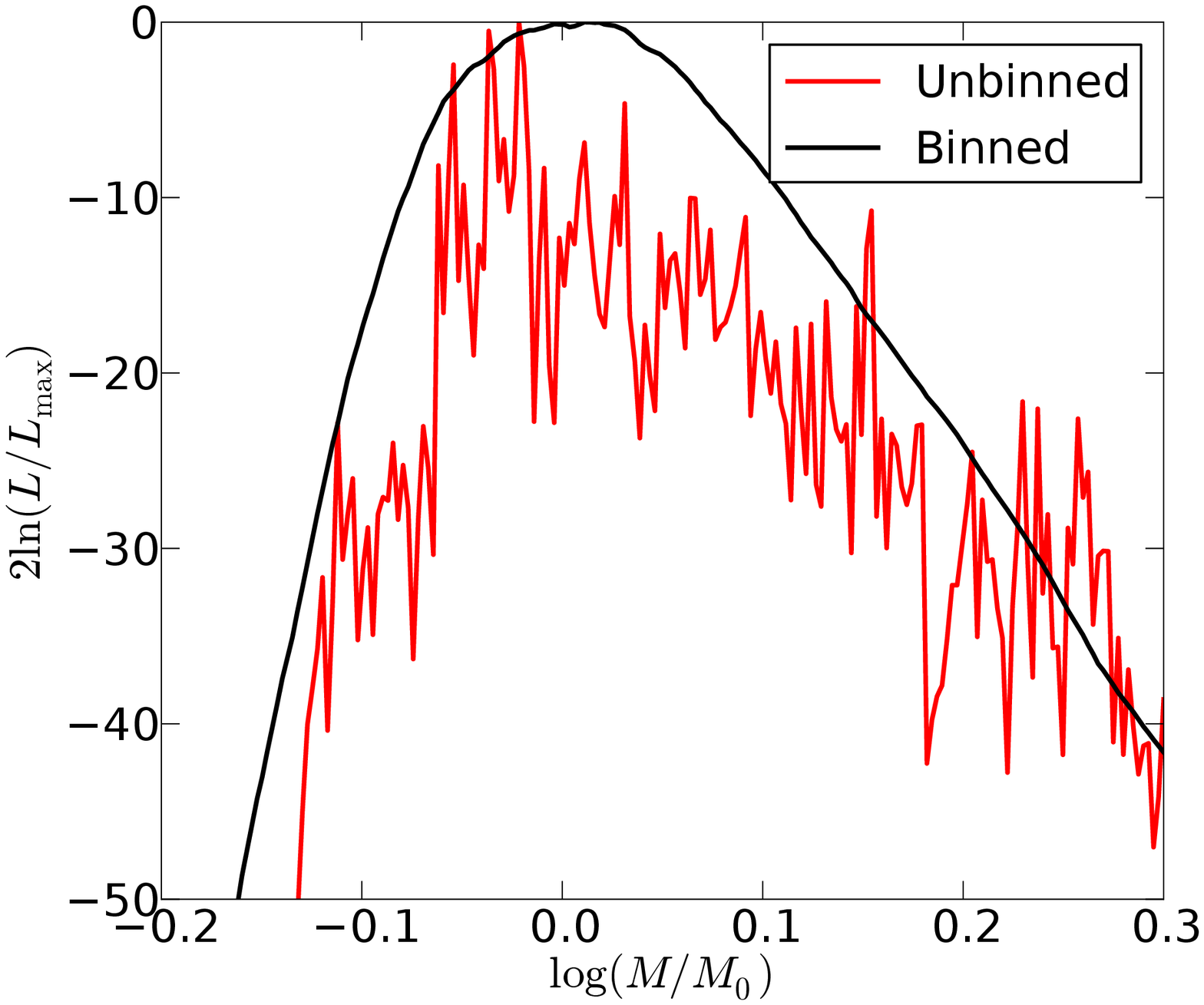}
 \caption{The radial likelihood function. We apply the unbinned
 (Eq.~\eqref{eq:like_unbin}, red spiky curve) and binned
 (Eq.~\eqref{eq:like_bin}, black smooth curve) likelihood estimator to the
 same sample of 1000 ideal tracer particles in an NFW halo. The
 concentration parameter is fixed to be the true value and the
 likelihood is calculated as a function of the mass parameter, $M$,
 normalized by the true mass, $M_0$. The likelihood ratio,
 $2\ln({\cal L}/{\cal L}_{\rm max})$, is plotted where 
  ${\cal L}_{\rm max}$ is the maximum
 likelihood value as the mass is varied. We adopt 30 bins logarithmically space between $1$ and $1000\,{\rm kpc}$ in radius in the binned case.}\label{fig:unbinned}
\end{figure}

A proper marginalization can be done if one knows the $P(E,L)$
distribution. If one introduces additional assumptions on
$P(E,L)$ \citep[e.g.,][]{Bovy}, then the method essentially
reduces to a $f(E,L)$ method, whose generality is limited by
the assumptions made. We would like to avoid any such assumptions in
order to avoid any potential bias in the inferred potential introduced through
them. Without prior knowledge of $P(E,L)$, we can
approximate it with the observed distribution 
\begin{equation}\label{eq_ELdiscrete}
 \frac{\D^2 P(E,L)}{\D E \D L}=\frac{1}{N}\sum_i \delta(E+K_i+\psi(r_i))\delta(L-L_i) .
\end{equation}
Now the marginalized distribution becomes:
\begin{align}
\D P(r)&=\int \D P(r|E,L)\, \frac{\D^2 P(E,L)}{\D E \D L} \, dE \, dL\\
&=\frac{1}{N}\sum_{i=1}^N \D P(r|E_i,L_i) .
\end{align}
This is the mixed (empirically marginalized) radial distribution, an
analogy to the marginalized theoretical radial distribution. We can
define a reciprocal probability of finding a particle at $r_i$ with the
orbital parameters corresponding to ($E_j,L_j$) as
\begin{align}
P_{ij}&=\frac{\D P(r_i|E_j,L_j)}{\D r_i}\label{eq:Pij}\\
&=\frac{1}{v_r(r_i,E_j,L_j)T_j} .
\end{align}
Then, the marginalized radial distribution becomes
\begin{equation}\label{eq:RadialProb}
\frac{\D P(r_i)}{\D r_i}=\frac{1}{N}\sum_{j=1}^N P_{ij} .
\end{equation}
This $P(r_i)$ is actually the posterior probability of finding a particle
at position $r_i$ given the orbital parameters of all the tracer
particles, $P(r_i|E_1,E_2,...E_N,L_1,L_2,...L_N)$. Now the likelihood
can be written as
\begin{equation}\label{eq:like_unbin}
{\cal L}=\prod_{i=1}^N \sum_{j=1}^N P_{ij} .
\end{equation}
However, the above likelihood function is very noisy. In
Fig.~\ref{fig:unbinned}, we show an example of this likelihood as a
function of the halo mass parameter for the ideal tracer. Overall, the
global shape of the likelihood function peaks around the true
parameter value. On the other hand, the likelihood function is not
at all smooth. Moreover, one does not get rid of this bumpiness by
zooming into a finer grid in $M$. The noisy behaviour originates from the
singularities in the PDF at peri and apo-centres, and from the
discreteness in the $(E,L)$ distribution, which we described as a sum of
$\delta$-functions. The noise in the likelihood can thus be regarded
as  Poisson noise. This Poisson noise prevents accurate inference of
parameter values. Some form of smoothing can
suppress the Poisson noise. For example, one could try to do kernel
interpolation in the $(E,L)$ distribution, to make it a continuous
rather than a discrete distribution~\citep[see
e.g.,][]{Bovy,Magorrian}. The simplest smoothing strategy would be to
bin the data. If we bin the data radially into $m$ bins, then the
binned version of the mixed radial likelihood can be written as
\begin{align}
 {\cal L}&=\prod_{i=1}^{m} \hat{n}_i^{n_i} \exp(-\hat{n}_i)\nonumber\\
 &=\exp\left(-\sum_i\hat{n}_i\right) \prod_{i=1}^{m} \hat{n}_i^{n_i}\nonumber\\
 &=\exp(-N) \prod_{i=1}^{m} \hat{n}_i^{n_i} , \label{eq:like_bin}
\end{align}
where $n_i$ is the number of particles in the $i$-th bin. We have
omitted the data-dependent constants in the above equation, and
$\sum_i \hat{n}_i=N$ due to the normalization of the PDF. The
predicted number of particles in the $i$-th bin is given by
\begin{equation}
\hat{n}_i= N\int_{r_{{\rm l},i}}^{r_{{\rm u},i}} \frac{\D P(r_i)}{\D r_i} \, d r_i,
\end{equation}
where $r_{{\rm l},i}$ and $r_{{\rm u},i}$ 
are the lower and upper bin edges, and
$P(r_i)$ is given by equation~\eqref{eq:RadialProb}. The binned
likelihood curve is also shown in Fig.~\ref{fig:unbinned} for the same
sample. Clearly, the Poisson noise has been suppressed and the likelihood
function is now smooth and usable for parameter inference.

\begin{figure*}
 \myplottwo{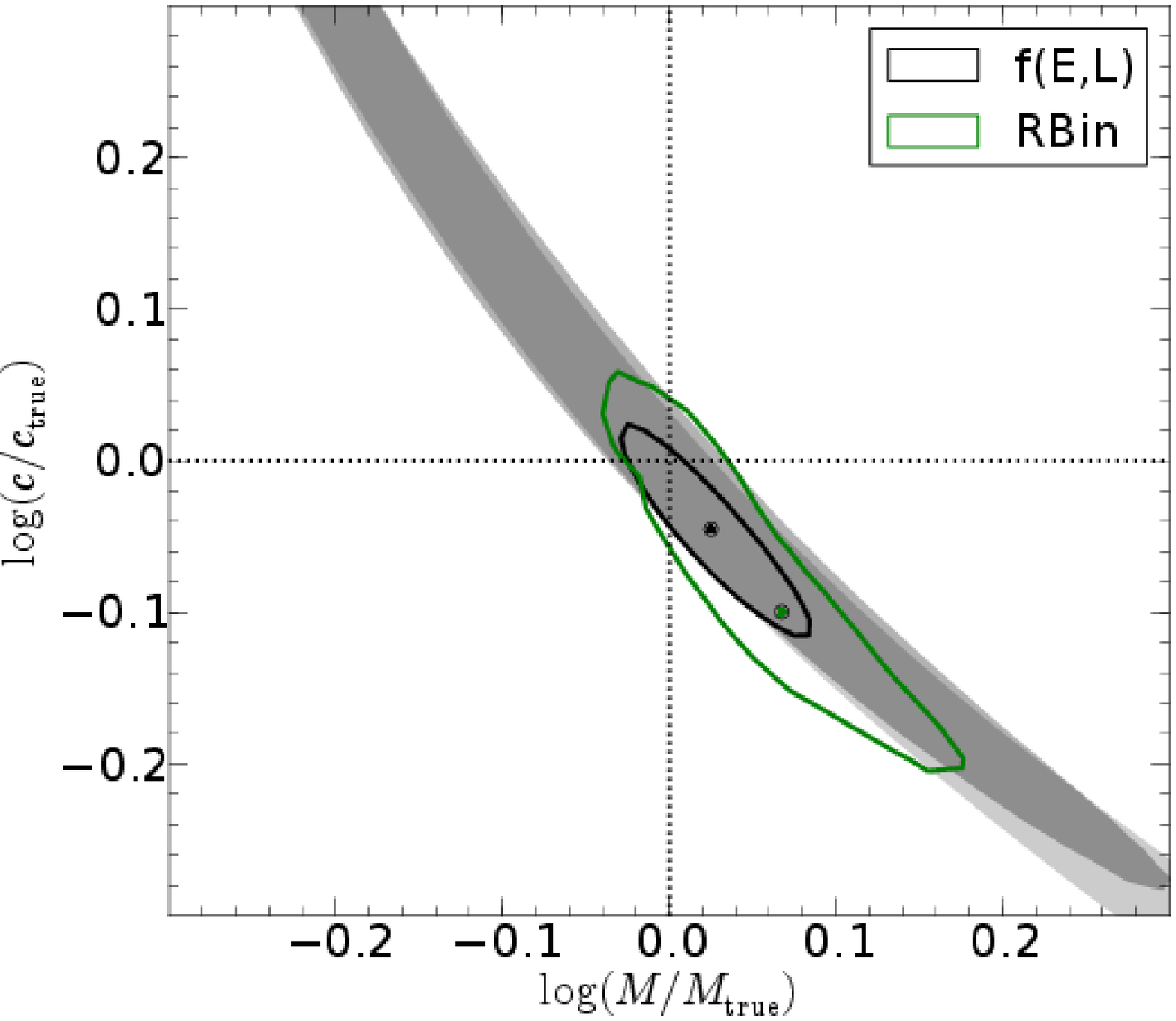}{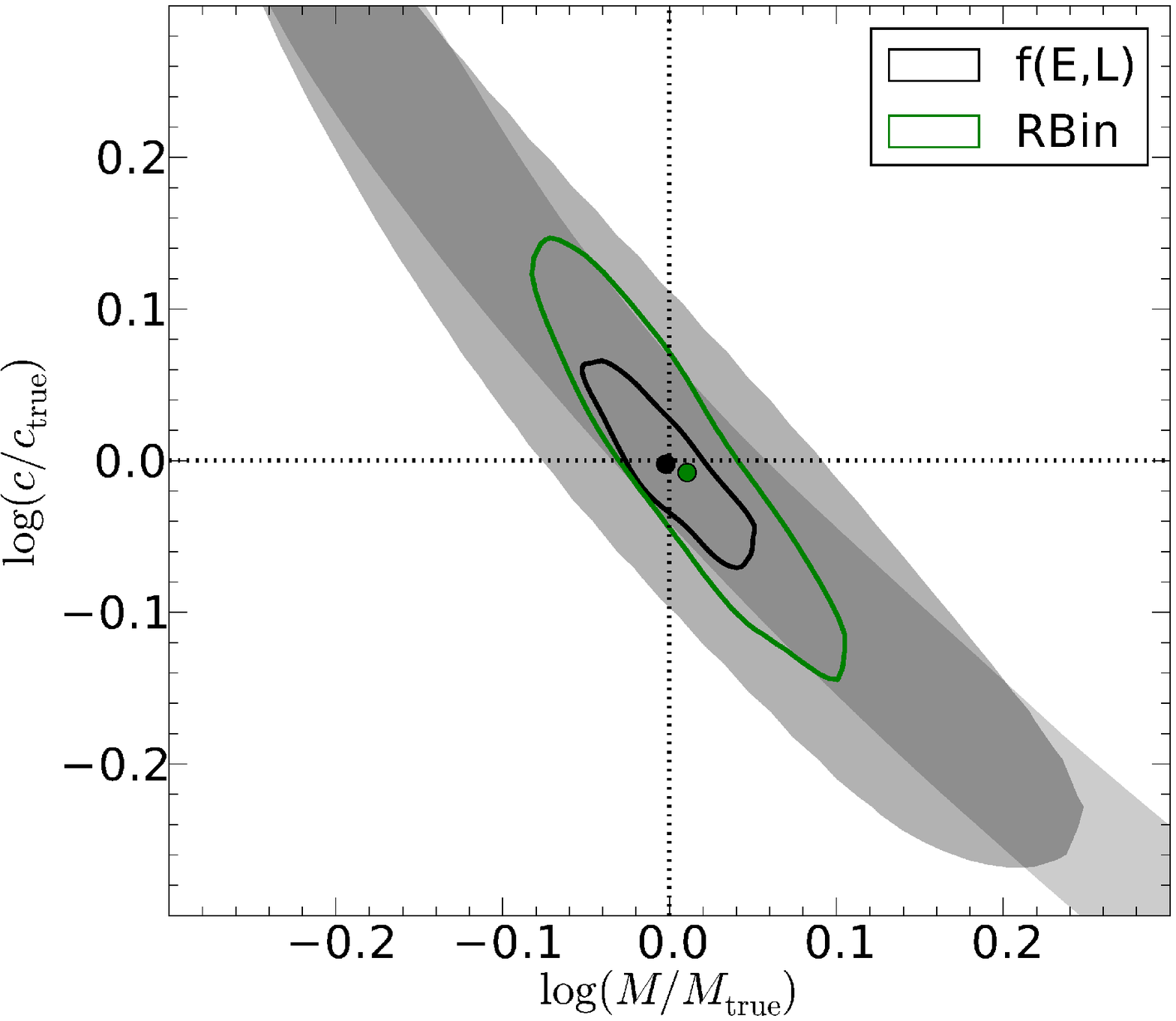}
 \caption{The $1\sigma$ confidence region for the different
 estimators. In both panels, the dark and light shaded regions are the
 confidence regions from the AD and mean-phase estimators
 respectively, while the green
 line marks the confidence region from the radial likelihood
 estimator (RBin), and the black line marks that from the $f(E,L)$
 estimator. In the left panel, the confidence regions are inferred
 from a single sample of 1000 particles. The blue and black points
 correspond to the best-fit parameters of the radial likelihood and
 $f(E,L)$ estimators. In the right panel, the confidence regions
 represent the 68.3\% most probable region of the best-fitting
 parameters, according to the distribution of best-fitting parameters
 from 750 independent samples. The blue and black points correspond
 to the average parameters from the radial likelihood and $f(E,L)$
 estimators respectively. Note that since the AD and mean phase fits
 are sensitive to the initial guess of the parameter values due to
 their strong degeneracy, we have randomly picked the initial values
 when fitting each sample. The $f(E,L)$ and RBin fits are independent
 of the initial parameters. The samples used here are generated
 according to the $f(E,L)$ DF with true parameters $(M_0,c_0)$; hence
 the $f(E,L)$ fit is, by construction, the best constraint one can
 obtain.}\label{fig:RBinLike}
\end{figure*}

In Fig.~\ref{fig:RBinLike} we apply the binned radial likelihood
estimator to the ideal tracers. In the left panel, the three
estimators are applied to the single realization of 1000 particles
used in Fig.~\ref{fig:MinDist}. The degeneracy we have seen in the
phase angle estimators is broken in the radial likelihood
estimator. The estimators are applied to many (750) independent
realizations of the same $f(E,L)$ distribution, and the distribution
of the best-fit parameters are plotted in the right panel of
Fig.~\ref{fig:RBinLike}. This test shows the estimator to be unbiased 
when averaged over many realizations.
For comparison, we also show the result
of a likelihood analysis using the full $f(E,L)$ distribution as in
\citet{Wang}. Note that since the data are generated with this exact
DF, the $f(E,L)$ likelihood applied to these ideal
tracers represent the best constraint one can get from any likelihood
inference.  


The radial likelihood estimator gives wider but still comparable
confidence intervals as the full $f(E,L)$ estimator. Note that we have not assumed anything about the $(E,L)$ distribution
of the tracers and that the mock catalogue is used blindly. The above test
is a general demonstration that our method will work for any
steady-state tracers in a static spherical potential. 
Compared to the perfect $f(E,L)$ method, where the adopted
$f(E,L)$ exactly matches the form of the unknown underlying distribution
function, the confidence interval is wider but, in practice, the
$f(E,L)$ method will be biased if the tracer follows a
$(E,L)$ distribution other than the specific $f(E,L)$ model adopted. With a
small increase in noise, our radial likelihood method gains generality.
Compared with the minimum distance estimators, the radial
likelihood estimator breaks the degeneracy in the mass concentration
parameters. Since the likelihood can be interpreted as the conditional
probability of the data given the model, in principle it can also be adopted in an Bayesian analysis.

We now make a few comments on the practical application of the binned
likelihood estimator. Since the purpose of the binning is purely
to suppress shot noise, a larger number of bins is generally better,
as long as it is not too noisy. On the other hand, when the likelihood
contours appear too irregular, one should try reducing the number of
radial bins to ensure the irregularities are not caused by shot
noise. In our analysis, we have adopted 30 logarithmic bins for an
ideal sample of 1000 particles, and 50 bins for $10^6$ particles in a
realistic halo (paper~II), although as few as 5 bins could
still work. Due to the singularity of the $1/v$ integrand at orbital
boundaries, it is expensive to achieve high accuracy for the phase
calculations. As a result, it is difficult to use algorithms involving
numerical derivatives for the optimization of the likelihood
values. Instead, we adopt the Nelder-Mead simplex minimizer to search
for the maximum of the likelihood. 
\section{Reconstructing the potential profile: the phase-mark non-parametric method}\label{sec:mark} 
\begin{figure*}
 \myplottwo{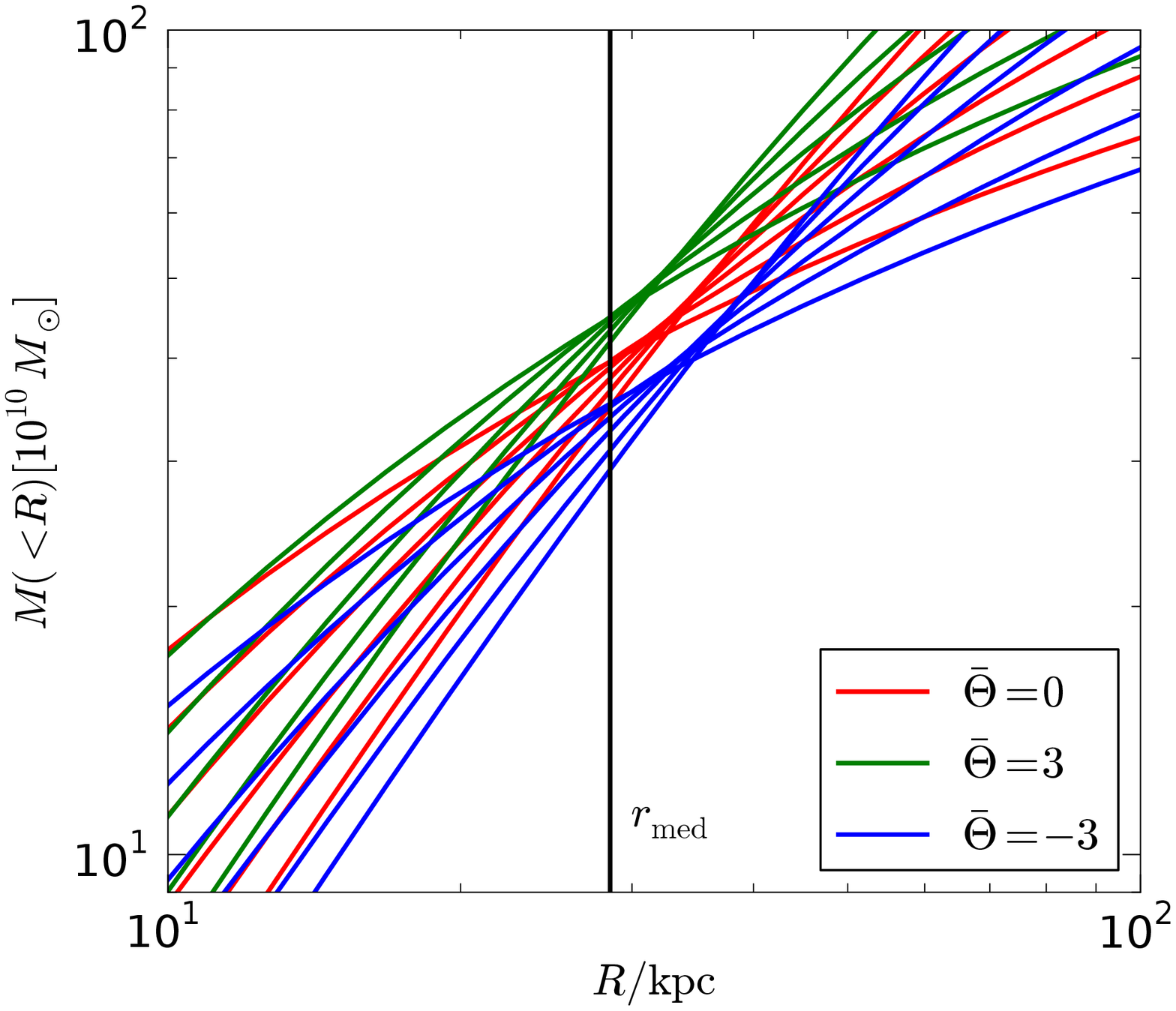}{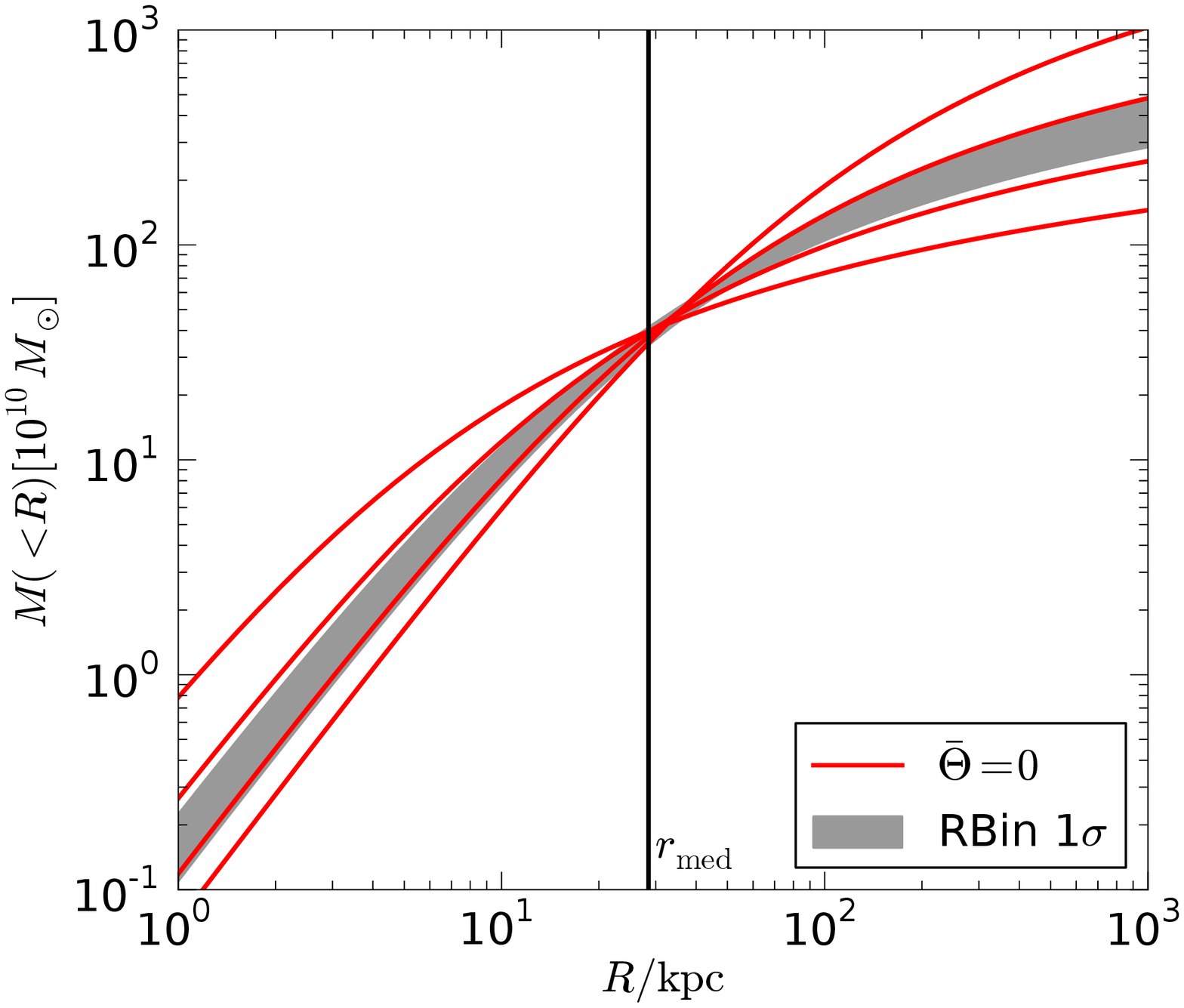}
 \caption{Constraints on the mass profile from different methods. In
 the left panel, we plot the predicted mass profiles adopting
 parameters lying on the $\bar{\Theta}=0$ and~$\pm3$ lines in the parameter
 space in Fig.~\ref{fig:MinDist} ($\bar{\Theta}=3, 0, -3$ from top to bottom near the median radius). The vertical line marks the median radius (i.e., the half mass
 radius) of the tracer. In the right panel, the $\bar{\Theta}=0$
 profiles are compared, over a wider radial range, with those 
adopting parameters on the $1\sigma$
 contour of the binned radial likelihood estimator. The span of the
 latter, that is, the $1\sigma$ prediction bounds of the likelihood
 estimator, are marked by the shaded region. Note that since the constant
 $\bar{\Theta}$ lines are not closed in the parameter space (e.g.,
 Fig.~\ref{fig:MinDist}), there could be many mass profiles with
 shapes far more different from those plotted in this figure that still
 share the same $\bar{\Theta}$. In other words, the shape variation at
 the same $\bar{\Theta}$ could be much larger than 
 plotted.}\label{fig:MassProf}
\end{figure*}

\begin{figure*}
\myplottwo{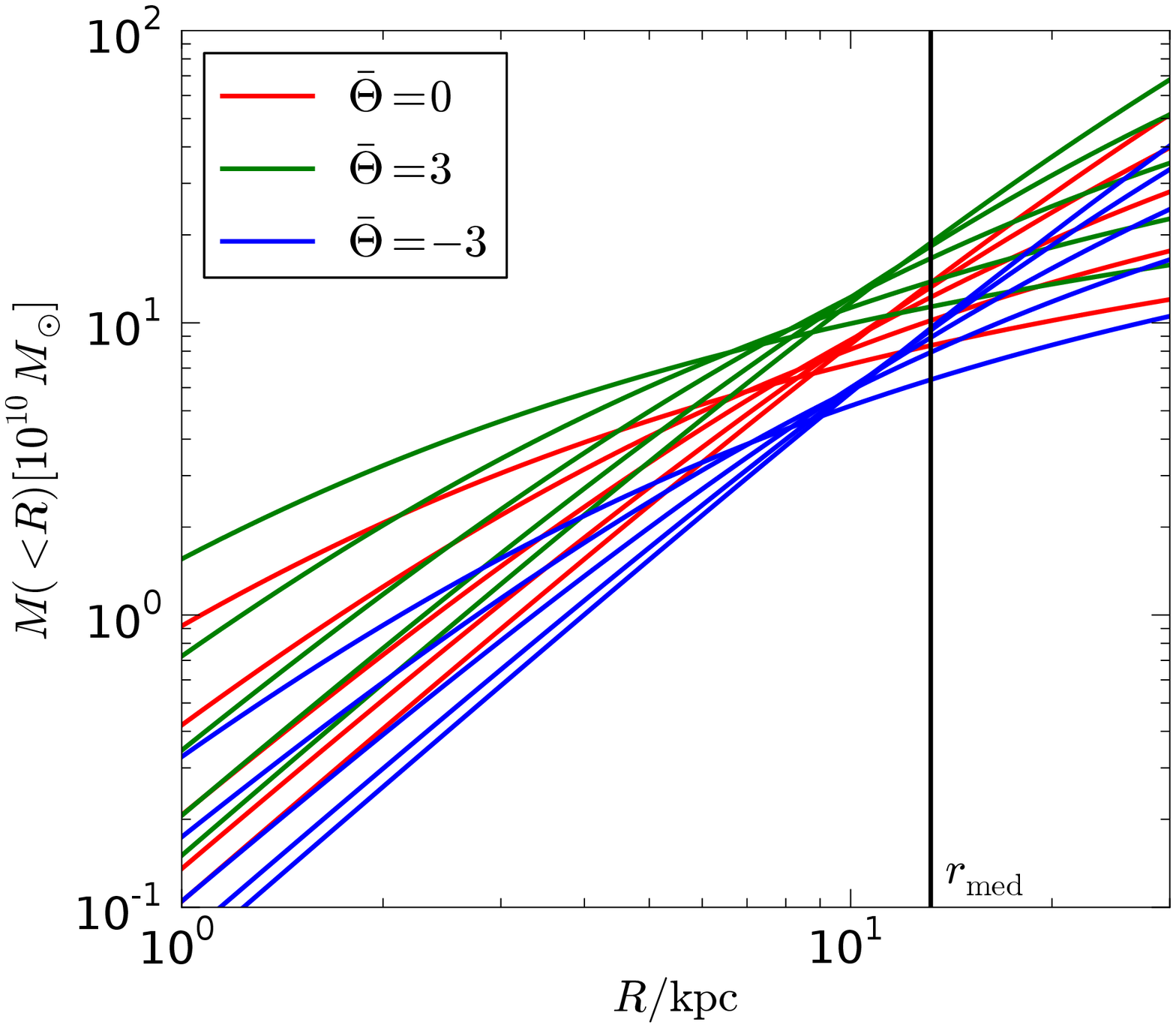}{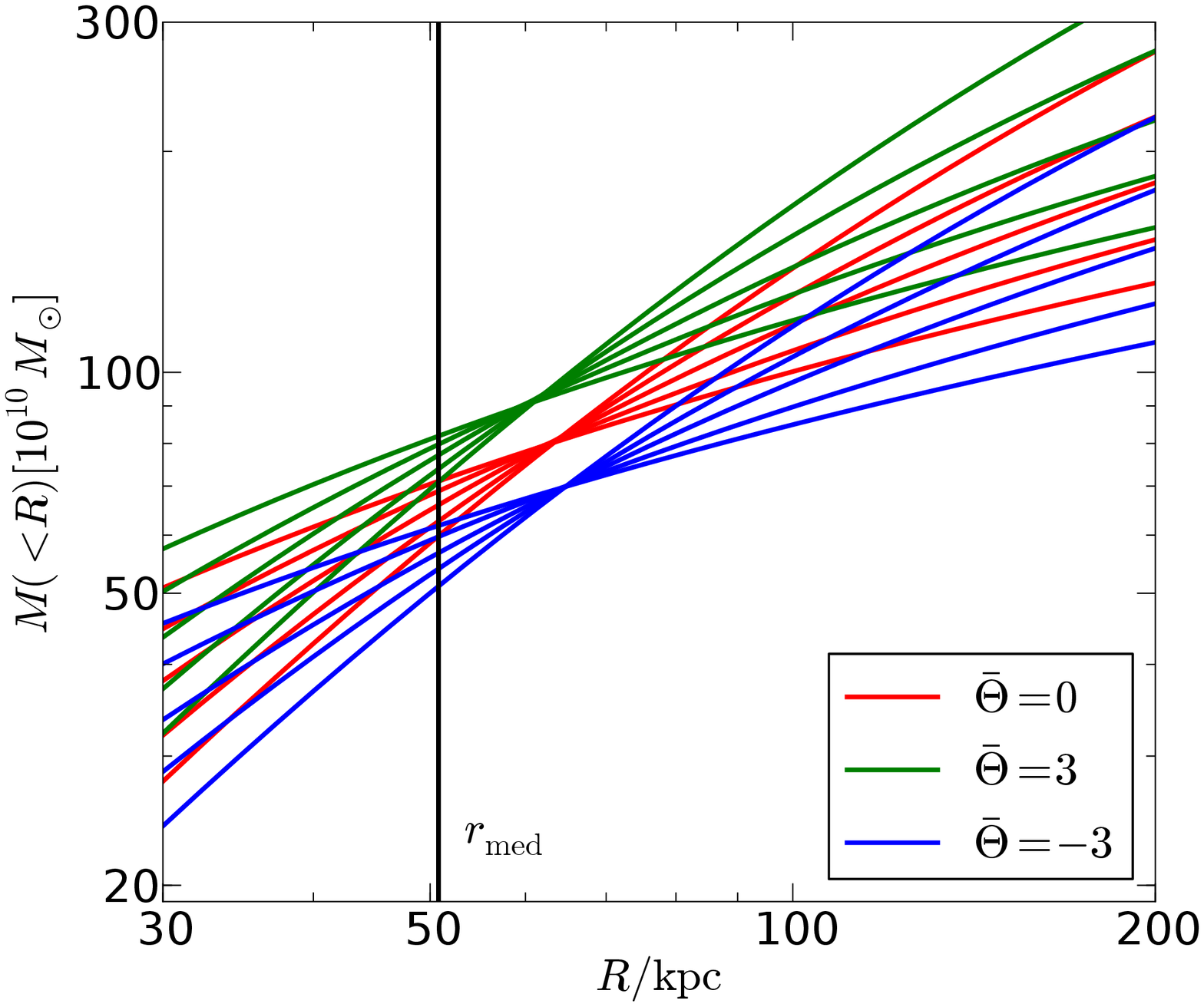} 
\caption{Mass profiles adopting parameters on constant $\bar{\Theta}$ lines, demonstrating the characteristic radius of the tracers. These are the same as the left panel of Fig.~\ref{fig:MassProf} but using two tracer samples with different radial cuts. The left panel uses a sample from $1-30~{\rm kpc}$, while the right panel uses a sample from $30-1000~{\rm kpc}$. The vertical black lines mark the half mass (i.e., median) radius of each sample. The sample sizes are both 1000 particles and are selected by applying only radial cuts to the parent samples constructed in Section~\ref{sec:data}.}\label{fig:MassProf2}
\end{figure*}

\begin{figure*}
 \myplottwo{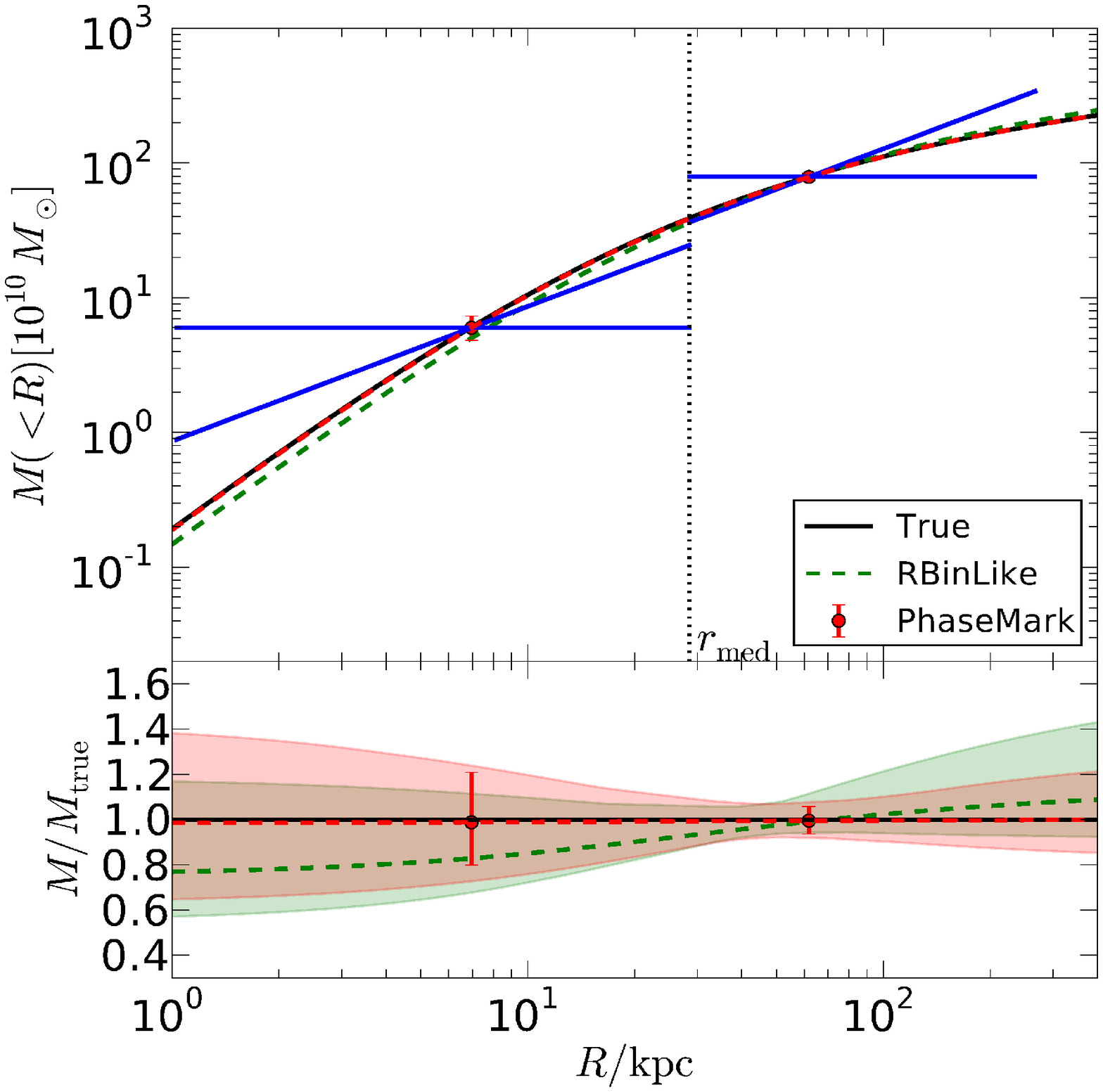}{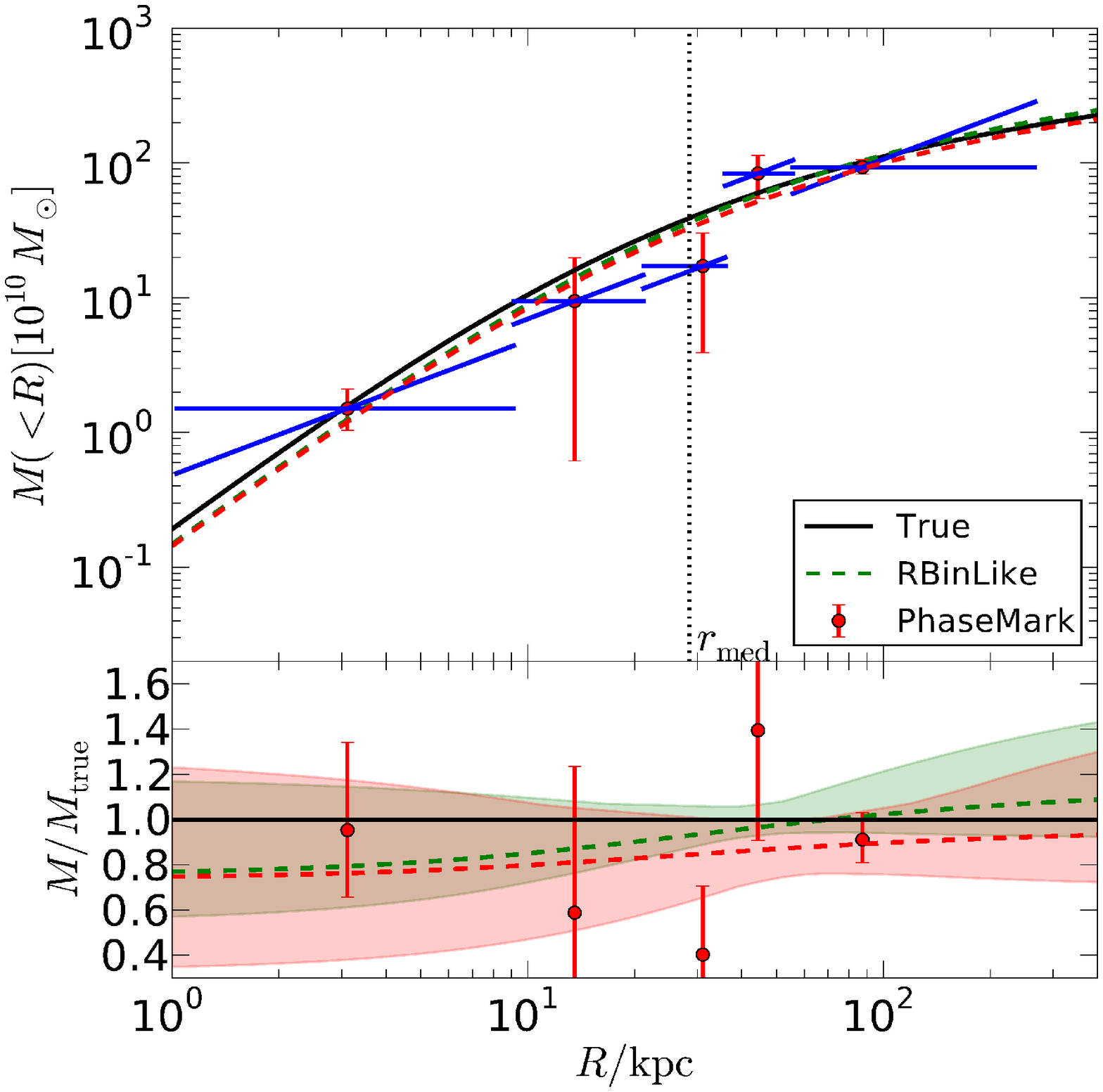}
 \caption{The mass profile constrained with the phase-mark method. The left and right panels are the same except that they have different number of radial bins. The bins are defined to have equal numbers of tracer particles, by subdividing a parent sample of 1000 particles. In each upper panel, the vertical dotted line marks the median radius of the full sample of tracers; the black solid line shows the true mass profile of the halo; the green dashed line is the best-fit profile using the radial likelihood method; the blue solid lines are the best-fit point-mass and isothermal profiles in each radial range. The point where the two blue lines cross (marked by points with error-bars) gives the characteristic mass in each bin. The errorbars are the uncertainty in the fitted point-mass parameter. The red dashed line is the best-fit NFW profile to the characteristic mass points. The bottom panels are the corresponding mass profiles divided by the true profile. The shaded region shows the $1$-$\sigma$ uncertainty on the fitted profiles from the radial likelihood (green) and from fitting NFW profiles to the characteristic masses (red).}\label{fig:PhaseTicker}
\end{figure*}

\subsection{Towards understanding the parameter degeneracy}
The strong parameter degeneracy with the minimum distance estimators
is easy to understand when we examine the constraints on the mass profile of the
halo, as shown in Fig.~\ref{fig:MassProf}. Parameters yielding the
same mean phase deviation, $\bar{\Theta}$, all predict the same mass $M(<R_{\rm c})=M_{\rm c}$ inside a
characteristic radius, $R_{\rm c}$, of the tracer, which is close to the half-mass radius of the tracer. 
Different $\bar{\Theta}$ values correspond to different $M(<R_{\rm c})$, with a positive correlation between the
two. In other words, the mean phase of the tracer is an estimator of
the gravitational force $G M(<R_{\rm c})/R_{\rm c}^2$ or circular velocity around the characteristic radius of the
tracer population. On the other hand, this estimator barely constrains the
gravity elsewhere, leaving the shape of the mass profile
unconstrained. Parameters leading to the same $M(<R_{\rm c})$ but different
shapes in the mass profile are thus indistinguishable by the mean
phase estimator, resulting in the parameter degeneracy in
Fig.~\ref{fig:MinDist}. The radial likelihood estimator breaks this
degeneracy by its ability to also constrain the shape of the profile, 
as illustrated in the right panel of Fig.~\ref{fig:MassProf}. 

\begin{figure*}
 \myplottwo{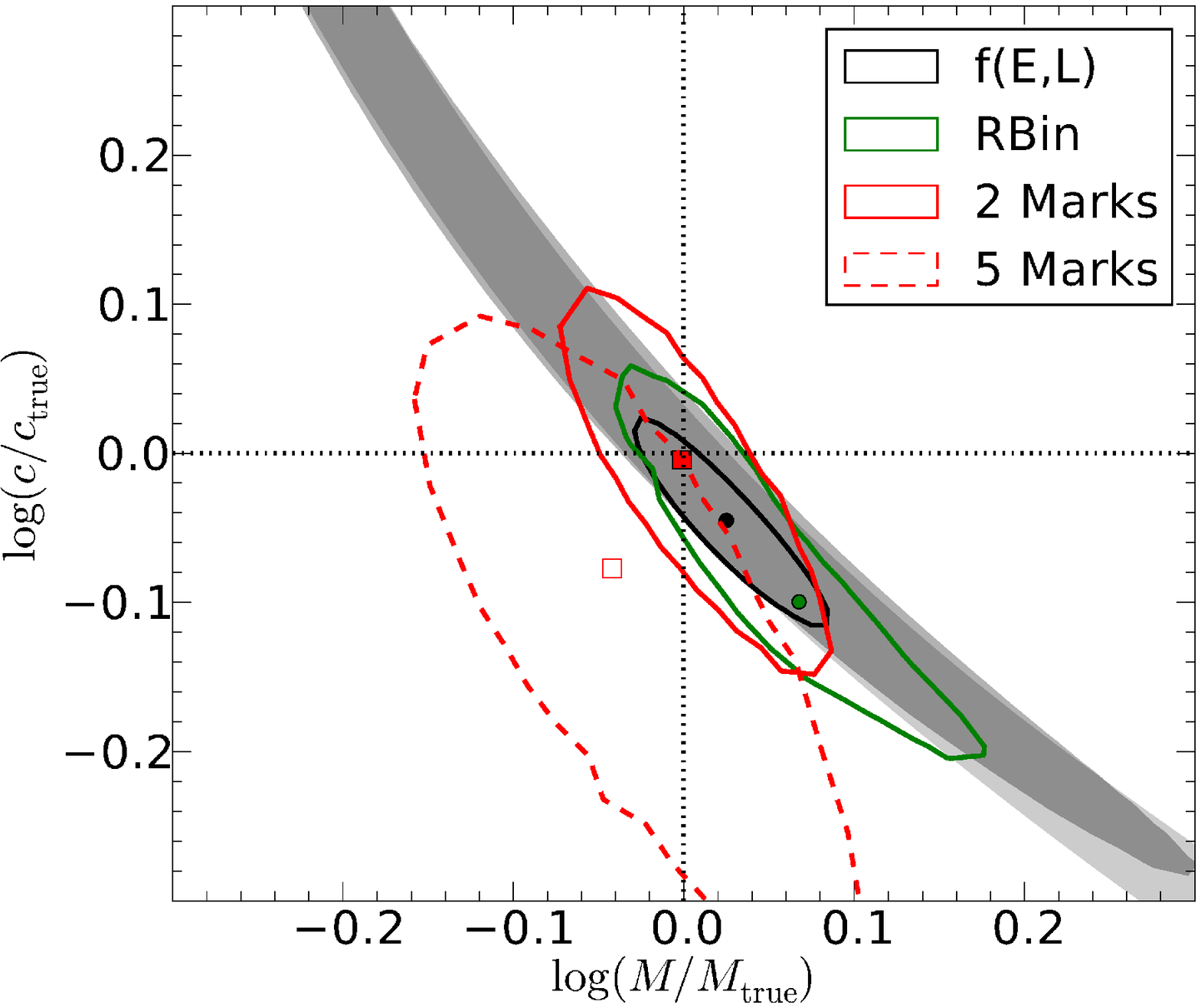}{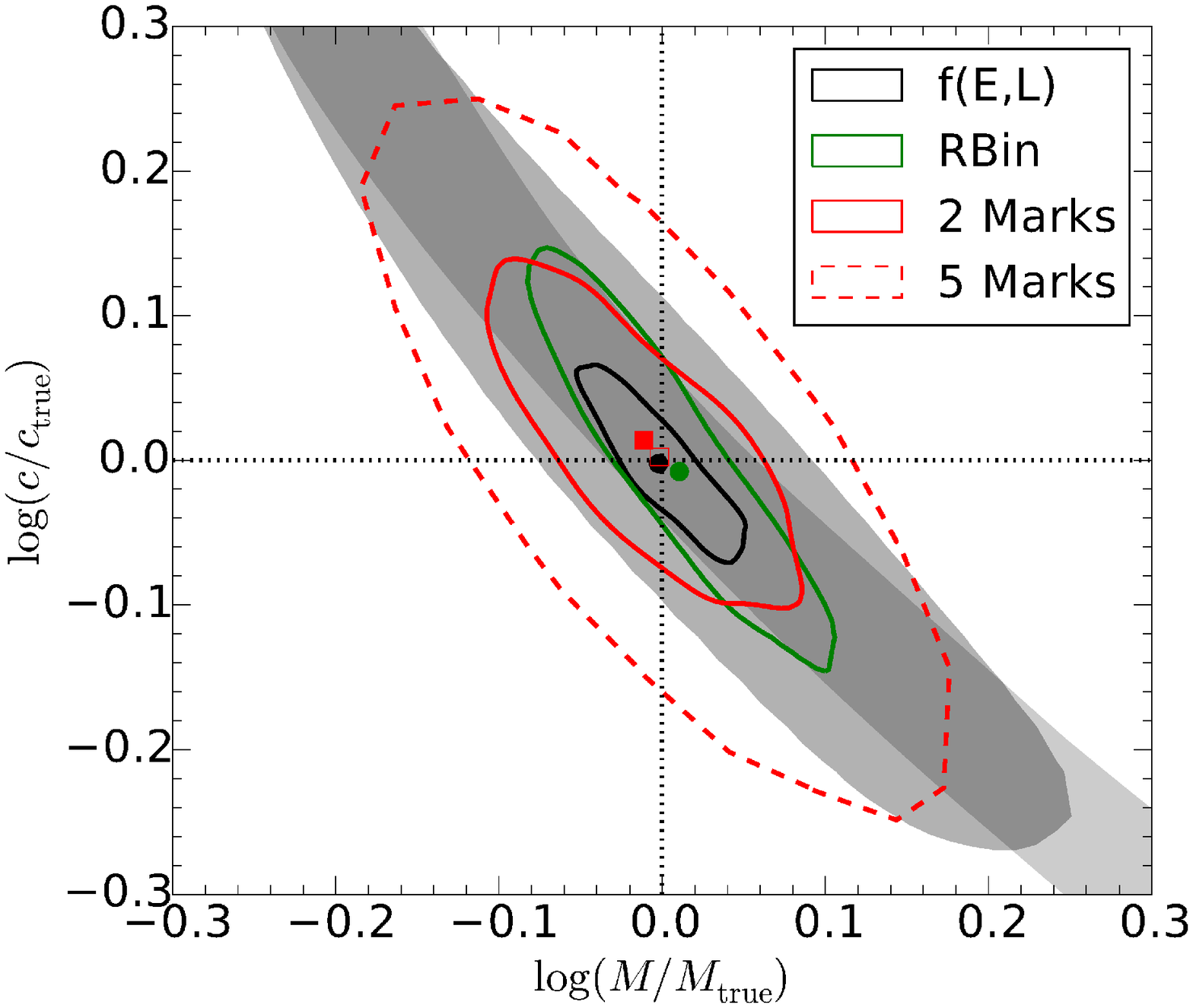}
 \caption{Same as Fig.~\ref{fig:RBinLike}, but also showing the constraints from fitting NFW profiles to the phase-marks. In the left panel, the red solid and red dashed contours show the $1$-$\sigma$ confidence regions obtained by dividing a single sample of 1000 particles into two and five radial bins respectively. The red filled and open squares in the centre of each contour show the corresponding best-fit parameters. In the right panel, the contours show the 68.3\% most probable region of the best-fitting parameters, according to their distribution obtained from many (750) independent samples of 1000 particles each. The points in the centre show the median best-fitting parameters. Again the result for the phase-mark with two and five bins are shown in red solid and dashed contours respectively, while the filled and open squares show the corresponding median parameters.}\label{fig:TickerContour}
\end{figure*}

The positive correlation between the mean phase $\bar{\Theta}$ and the
characteristic mass, $M(<R_{\rm c})$, can be understood qualitatively. For
profiles with the same shape (on a logscale), a higher $M(<R_{\rm c})$ leads
to deeper potential everywhere. For a particle with a given position
and velocity, a deeper potential of the same shape will shift both its peri
and apocentres closer to the centre of the halo. As a result, the
current location of the particle relative to its peri and apocentres
is shifted outward, increasing its phase angle $\theta$. The mean
phase of all particles thus increases with the characteristic mass.

The location of the characteristic radius determines the shape of the mean
phase line in parameter space. To demonstrate this,
instead of working in the $(M,c)$ parameter space, it is more
convenient to work in the $(M_{\rm s},r_{\rm s})$ space, where $M_{\rm s}=4\pi\rho_{\rm s}
r_{\rm s}^3$. Suppose the constant $\bar{\Theta}$ lines are described by
$M_{\rm s}=f_{\bar{\Theta}}(r_{\rm s})$, then we have
$M(<r)=f_{\bar{\Theta}}(r_{\rm s})[\ln(1+r/r_{\rm s})-(r/r_{\rm s})/(1+r/r_{\rm s})]$. Now the
characteristic radius $r$ at which the mass does not vary with the halo
parameter $r_{\rm s}$ is given by
\begin{equation}
 \frac{\partial M(<r,r_{\rm s})}{\partial r_{\rm s}}=0,\label{eq:Rchar}
\end{equation} whose solution $r=R_{\rm c}$ depends on the contour line function $f_{\bar{\Theta}}(r_{\rm s})$. 
So the functional form of the contour line determines $R_{\rm c}$, and vice versa.

It is worth clarifying that the characteristic radius, hence also the shape
of the degeneracy curve, is determined by the distribution of
the tracer and is not an intrinsic property of the halo. We
demonstrate this in Fig.~\ref{fig:MassProf2}, where the characteristic
radii of two ideal tracers with different radial ranges are shown. The
two tracers are constructed by sampling from $1-30$ and $30-1000$~kpc
respectively from a parent population whose half mass radius is
roughly $30$~kpc. Clearly, the two samples have different characteristic
radii, which are both close to their own half-mass radii, while the
haloes hosting the two samples are identical. Correspondingly, we
have checked that the mean phase contours have different slopes in
$\log(M)-\log(c)$ space from that shown in Fig.~\ref{fig:MinDist}. 

The existence of a best-constrained mass despite the parameter degeneracy is broadly in line with empirical results on the robustness of the mass constraint inside the half-light radius from Jeans equation modelling of dwarf galaxies~\citep{Walker09,Wolf10}. The existence of such a characteristic point in the Jeans analysis is further proved by \citet{Wolf10}. However, their results are concerned with the insensitivity of the mass estimate to the velocity anisotropy parameter of the tracers, which has to be fitted or assumed because only line-of-sight velocities are available in these studies. On the other hand, our parameter degeneracy arises from solving the mean-phase equation using the full 6D data. Anisotropy is not a parameter in our model at all. 

Another closely related result to ours is presented by \citet[AE11 hereafter]{AE11}, who studied the underlying potential of dwarf spheroidals assuming a lowered isothermal DF. They found that the structural parameters $(\rho_s, r_s)$ (or $(\rho_0, r_0)$ in the original notation of AE11) of NFW-like haloes are constrained by the observed $(R_{\rm h},\sigma_0)$ of each dwarf spheroidal to follow a certain relation, $\rho_s(r_s)$, where $R_{\rm h}$ is the projected half-light radius and $\sigma_0$ is the line-of-sight velocity dispersion at the centre of the dwarf spheroidal. This resulted in a best-constrained mass near a common characteristic radius $R_{\rm c}=1.7 R_{\rm h}$ for almost all the dwarf spheroidals. As we explained above (see Equation \ref{eq:Rchar}), the existence of this characteristic point can be understood because the two parameter halo profile is reduced to a one parameter family due to the constraint of the problem. In AE11, the constraint comes from matching the observed $(R_{\rm h}, \sigma_0)$ of each system. By contrast, in our case a constrained relation, $\rho_s(r_s)$, or equivalently, $M(c)$, is determined by solving $\bar{\Theta}=0$ (see Fig.~\ref{fig:MinDist}). Note that it is not expected that an arbitrary constraint (for example, $r_s(\rho_s)=\rm{Const}$) would always result in a best-constrained mass in the mass profile, so the similarity between the results by AE11 and ours is intriguing. Despite the apparent similarity, our result is purely theoretical, while that of AE11 is empirically driven by the observed quantities, $(R_h,\sigma_0)$. Our finding is expected to apply to steady state tracers in general, not only to those described by the lowered isothermal DF or to the observed dwarf spheroidals studied in AE11. In our general case, the characteristic radius is not a constant value in units of the median radius, as is evident in Fig.~\ref{fig:MassProf2}. We checked that the same is true (i.e., the scale is not universal) in units of the projected half-mass radius. By contrast, the common characteristic radius in AE11 is likely to arise from some common properties shared by the dwarf spheroidals studied there, namely the tight correlation between $(R_h,\sigma_0)$. 

\subsection{The phase-mark method}
The experiment in Fig.~\ref{fig:MassProf2} also suggests a way of breaking the degeneracy in the mean-phase estimator, by applying it to two or more subsamples split in radius. In this way the shape of the mass profile can be constrained as well, since different subsamples constrain the characteristic mass, $M_{\rm c}$, at different radius, $R_{\rm c}$. For parametric fits, the degeneracy lines in Fig.~\ref{fig:MinDist} would have different slopes for subsamples with different $R_{\rm c}$, so they could jointly determine a unique best-fit parameter set. Even better than that, it is possible to reconstruct the mass profile non-parametrically, thanks to the insensitivity of the mean-phase constraint on the shape of the mass profile. The fact that the mean phase only depends on the characteristic mass point means one can start from a profile of an arbitrary shape and still obtain the correct characteristic mass, by requiring the profile to produce the correct mean phase. As a result, it is not necessary to know the functional form of the true profile in order to constrain $(R_{\rm c}, M_{\rm c})$. By applying the mean-phase constraint $\bar{\Theta}=0$ twice to two different single-parameter profiles and looking for the point where they intersect, one can simultaneously obtain both the characteristic radius and the characteristic mass. Note that only having $M_{\rm c}$ is not enough, since $R_{\rm c}$ is unknown even though it is close to the tracer half-mass radius.

We demonstrate this in Fig.~\ref{fig:PhaseTicker}. In the left panel, we divide the tracer sample of 1000 particles studied in Fig.~\ref{fig:Degeneracy} into two sub-populations according to radius. Inside each bin, we fit two mass profiles with the mean-phase estimator: 1) point mass profile, $M(<R)=M_{\rm c}$ with parameter $M_{\rm c}$; 2) isothermal profile, $M(<R)=kR$, with parameter $k$. The mean phase constraint $\bar{\Theta}=0$ uniquely determines a best fit for each profile. As expected, they cross the true profile at the same point, which marks the characteristic point of that bin as ($R_{\rm c}=M_{\rm c}/k$, $M(<R_{\rm c})=M_{\rm c})$. We name this method the ``phase-mark''. Splitting the tracer into more bins, we can obtain finer constraints on the profile, as shown in the right hand panel. By doing this, we have reconstructed the true mass profile non-parametrically, without any assumption on the true profile. Such reconstructed mass profile becomes noisier when a larger number of bins is adopted (right panel), since each subsample becomes smaller. 

If desired, one can still fit a parametric function through the reconstructed profile, as shown by the red dashed line in each panel, with confidence regions on the fitted profile marked by the red shaded regions in the lower panels. After combining all the radial bins, the tightest constraint is still found near the half-mass radius of the full tracer sample. It is interesting to see that although a larger number of bins helps to obtain finer reconstruction of the profile, it does not lead to a better constrained profile after fitting. It appears that the constraining power using only two bins is close to that of the radial likelihood method. This is confirmed by the more direct comparisons made in Fig.~\ref{fig:TickerContour}. The confidence region of the phase-mark with two bins has a comparable size to that of the likelihood method. This means they are similarly efficient at making use of the dynamical information. Adopting finer bins in the phase-mark results in a looser constraint. This can be understood because each mark only exploits the local phase uniformity inside each radial bin, while the large scale variation from bin to bin is not taken into account, resulting in a leakage of information. A potential improvement would be to combine bins at different scales. However, it should be kept in mind that doing this will introduce correlations among the marks, making the error analysis difficult. In the right panel, the phase-mark with two bins is applied to many independent Monte-Carlo realizations of the same system as before, to show that the fit is statistically unbiased.

\section{Discussion}\label{sec:discussion}
\subsection{What is a tracer population?}\label{sec:discussion_tracer}
Dynamical modelling require the tracer sample
to be defined first, or subsamples to be selected from a parent
sample. Here we revisit the question of ``what is a tracer
population?''. 
Modelling the tracer population with a time-independent DF requires
the tracers to be in a steady state. For a spherically symmetric
potential, this requirement translates into a conditional radial
distribution $dN/dr \propto {1}/{|v_r(E,L,r)|}$ (necessary and
sufficient) given $E$ and $L$, or equivalently, all the particles have
completely uncorrelated radial phases. Once this condition
is satisfied, one can use the distribution of the sample to infer the
potential of the system. As a result, we can simply define a tracer as
any set of steady-state particles moving in the background potential.

To obtain a steady state subsample, the selection from the sample must
not distort this conditional radial distribution and must avoid
introducing artificial structure in the radial or angle
distribution. As long as this is  guaranteed, any selection in $E$
and $L$ is allowed. For example, one can select subregions of the
$E-L$ space, while keeping full or random sampling in $r$. For a
parent population not in equilibrium inside a static potential, a
steady-state subsample can still be selected by sampling according to
Eq~\eqref{eq:oPDFr}.

From this definition we also learn how to mix tracers with
weights. If tracer $i$ has a steady-state phase-space distribution
$f_i$, then the uniformly weighted population $w_i f_i$ is still a
tracer. Consequently, $N$ mixed tracers $\sum_{i=1}^N w_i f_i$ is
still a tracer, since equation~\eqref{eq:oPDFr} is satisfied for every
subtracer. As a result, when dealing with multi-component tracers
using the oPDF, we can either model them as a single population, or as
several populations separately. 

Obviously, subhalo particles are not steady-state tracers since they
are localized structures. Streams in general are also not steady
state tracers since they are usually characterized by correlated
phases.

\subsection{The optimal marginalization}\label{sec:marginalization}

\begin{figure}
\myplot{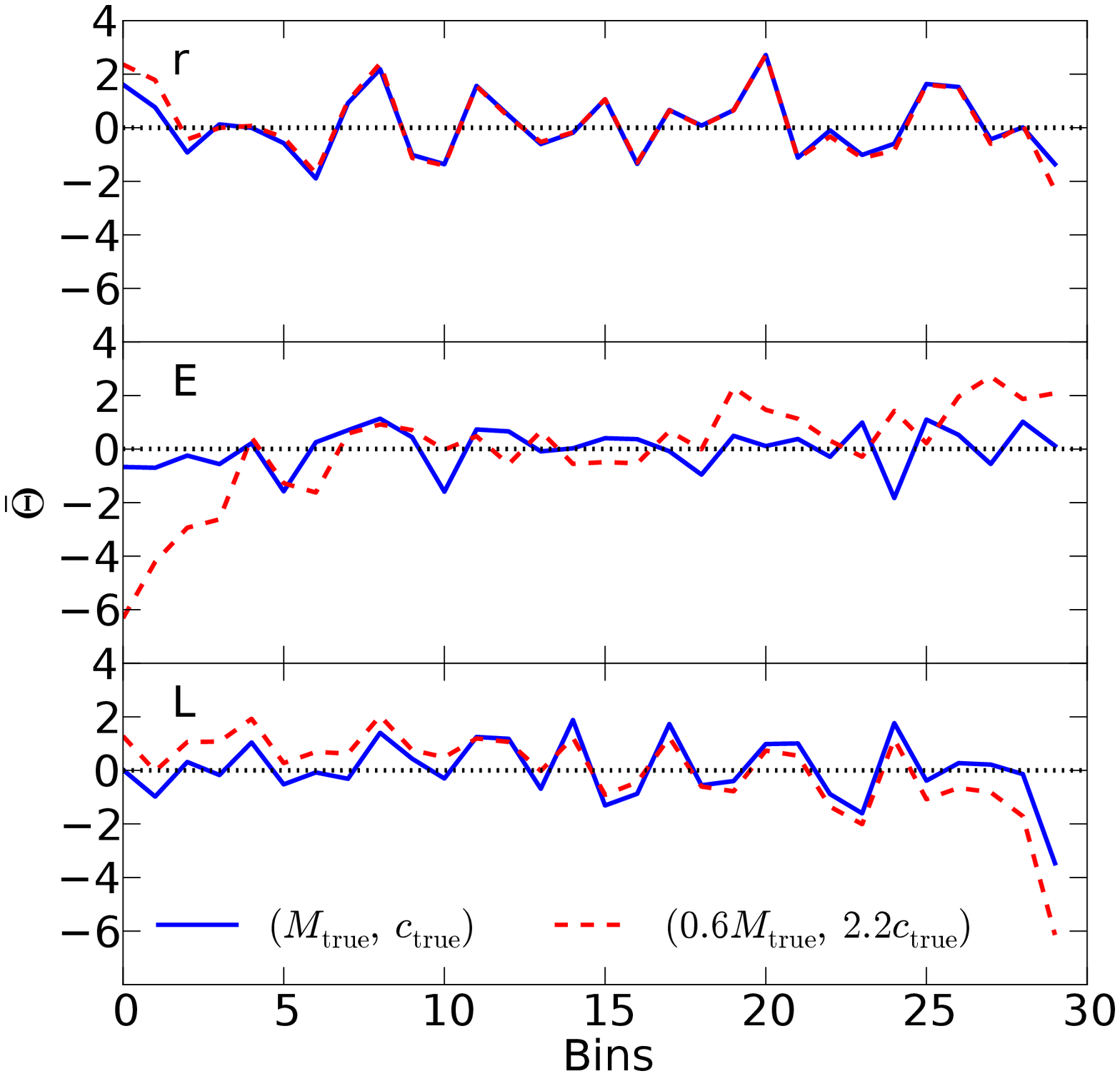}
\caption{Mean phase profile evaluated with the two degenerating parameter sets. From top to bottom we bin the same sample of 1000 tracer particles according to their $r,E,L$ coordinates respectively, with equal number of particles in each bin. The mean phase deviation, $\bar{\Theta}$, is evaluated inside each bin. Different coloured lines represent different haloes. The blue solid and red dashed lines are the profiles adopting two different potentials: one with the real parameter values ($M,c$) and the other with a parameter set ($0.6M,2.2c$) whose mean phase is degenerate with that of the true parameters. }\label{fig:MockTSprof}
\end{figure}

\begin{figure*}
\myplottwo{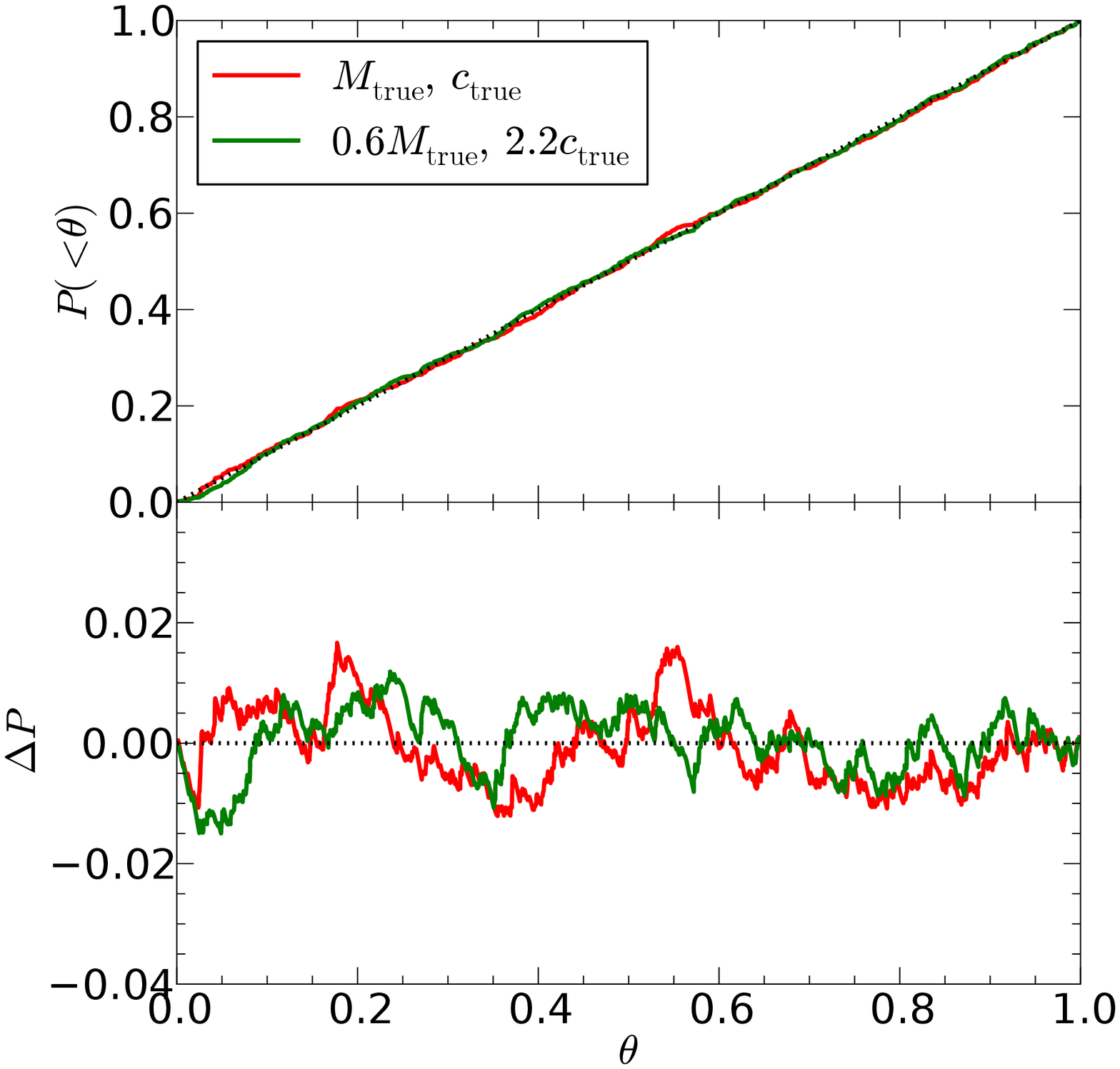}{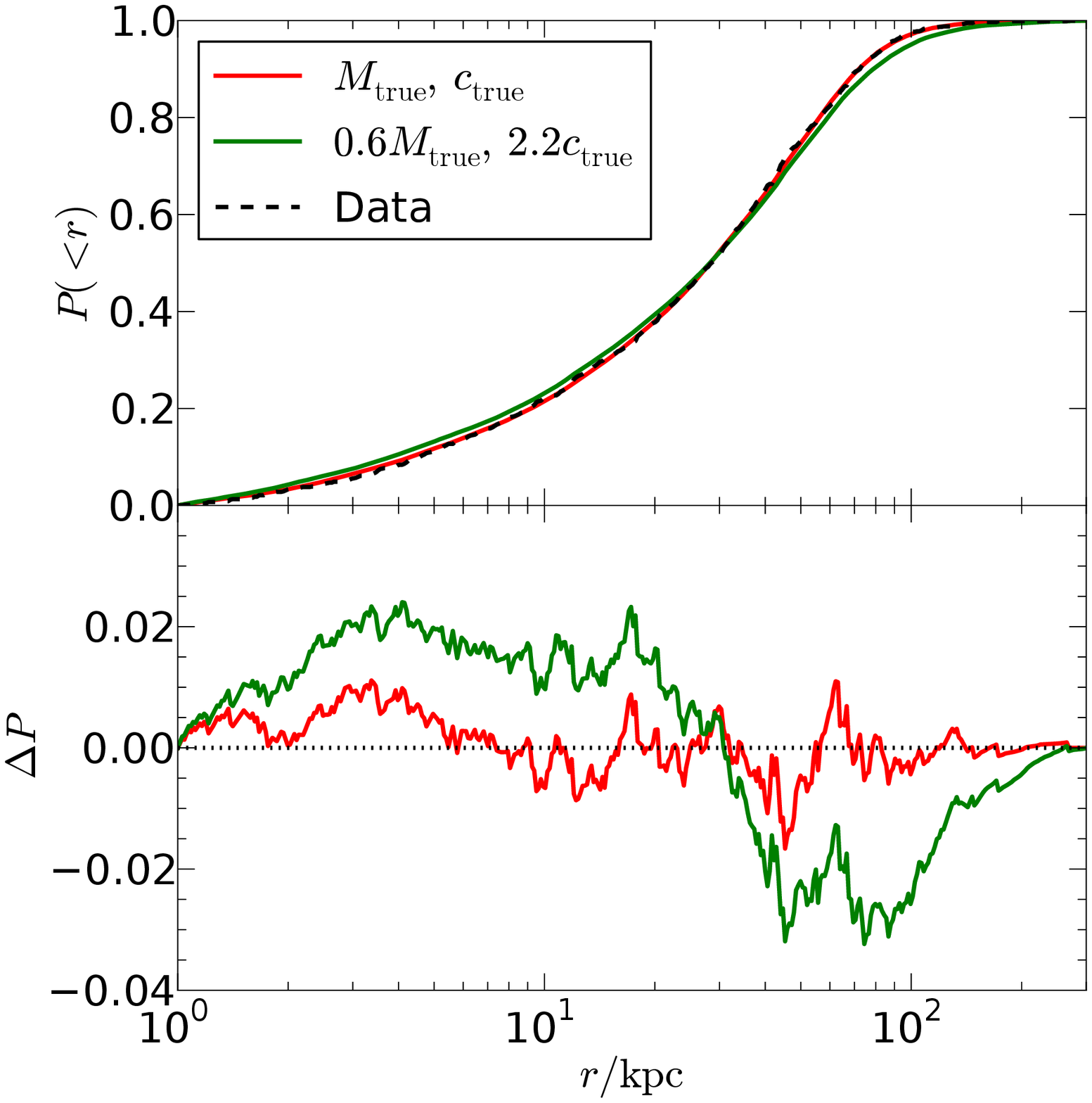}
\caption{Cumulative distributions in $\theta$ and $r$
spaces. Left: Cumulative distribution of $\theta$, at two sets of
parameter values. The red solid line corresponds to the true parameters,
while the green solid line corresponds to ($0.6M,2.2c$) which give the same mean phase. 
The bottom panel is the difference
between the model distribution and the cumulative distribution of a
uniform distribution.  Right: cumulative distribution of $r$. The
black dashed line is the distribution of the data, while the red and
green lines are for the true and alternatives parameters as in the
left panel.} \label{fig:Degeneracy}
\end{figure*}

It is tempting to ask why the radial likelihood estimator works better
than the phase estimators. Recall that the oPDF actually
specifies both the randomness of the phase angle and the independence
of such a distribution on the orbital parameters. That is, the
phase distribution is not only uniform in general, but also uniform inside any
$(E,L)$ bin. The oPDF also applies to any radial
range, so the uniformity is expected for any region in the $(r,E,L)$
space. In Fig.~\ref{fig:MockTSprof} we examine the mean phase in
subregions of the phase space. We divide the data into 30 equal-count
bins in each dimension, and measure the mean phase inside each bin for
a given potential. This mean phase value indicates the discrepancy of
the data inside this subregion with uniform phase distribution. We
do this for two mean phase degenerate points. For the true potential, $\bar{\Theta}$ is
consistent with 0 everywhere. For the degenerate parameter set, we start
to see a dependence on $(E,L)$, and $\bar{\Theta}$ is biased positively
or negatively at different places in the $(E,L)$ space, even though the
combined $\bar{\Theta}$ or the $\bar{\Theta}$ in $r$ space  would still
be close to 0. This test shows that there is still useful information
beyond the uniform $\theta$ distribution, namely its $(E,L)$
dependence. 



Since the minimum distance estimators do not examine the
$(E,L)$ dependence, they have effectively marginalized over the $(E,L)$
distribution of tracers. One can rewrite the definition of the phase
angle as
\begin{equation}
 \theta(r,E,L)=P(<r|E,L) .
\end{equation}
From this point of view, the marginalization is done by working in the 
cumulative probability space of the tracers. 
Although the radial likelihood method also marginalizes over the $(E,L)$
distribution, the marginalization combines the oPDF from different
$(E,L)$ orbits in a different (possibly optimal) way. This could result
in a marginalized DF that is more sensitive to the
discrepancies in the conditional distribution. In
Fig.~\ref{fig:Degeneracy} we see that the radial distribution is
indeed more sensitive, by examining the difference between empirical
and expected cumulative distributions in $\theta$ and $r$ space. 

\subsection{Connection to other methods}

Since $\D P(r,E,L)=\D P(r|E,L)\D P(E,L)$, the full phase-space distribution of
tracers breaks into two parts. The orbital PDF is determined by
dynamics, and reflects the underlying potential. The $P(E,L)$ part is
simply a characteristic of the tracer not necessarily related to
dynamics, and a sample with any form of $P(E,L)$ can be constructed
which is still a valid tracer population. Any DF
method has to make use of the $P(r|E,L)$ information in some way, but
how one deals with $P(E,L)$ is not crucial to the determination of the
potential.

\subsubsection{Comparison with the $f(E,L)$ DF method}\label{sec:DFconnection}

As we have already discussed in Section~\ref{sec:equiv_Jeans}, any
$f(E,L)$ DF function has to be consistent with the oPDF, while imposing
extra assumptions on the distribution of orbits. Because our method
only uses the conditional radial distribution, it is fully compatible
with the density profile inverted $f(E,L)$ method. At the same time, our
method has no extra assumptions and is applicable to more general
tracers. Also note that the density profile inversion involving a particular
parametrization of the potential can sometimes be quite challenging
~\citep[see, e.g.,][]{Wang}, and may not always be solvable
analytically. In contrast, the application of any potential function
in our oPDF method is always straightforward.

Since the oPDF is given in differential form, it does not care about
the radial limits of the system. One can apply the orbital PDF to data
within any radial range, e.g., from $r_{\rm min}$ to $r_{\rm max}$, since the
phase-space continuity equation holds within any radial range. When
radial cuts are imposed, we only need to replace the orbital limits
$r_{\rm a}$ with $\max(r_{\rm a}, r_{\rm min})$ and $r_{\rm p}$ with
$\min(r_{\rm p}, r_{\rm max})$. 
In this case, the data only care about the variation of
the potential within the same radial range. For the same reason, the
zero point of the potential, the extension of the halo or tracer
density profile outside the data window, or the boundary of the halo
is irrelevant in our method. By contrast, in the $f(E,L)$ method, the
DF, $f(E,L)$, has to satisfy the radial constraint $\rho(r)=\int
f(E,L)d^3v$ at all $r$ by definition. Any change in $\rho(r)$ at any
radius will require an adjustment in the proposed DF. As a
result, one has to include the full radial range of each orbit in its
density profile inversion. This requires a full description of both the
potential and the tracer density over the full radial range,
introducing a dependence on quantities outside the data window. Such
dependence, in turn, requires one to parametrize the tracer and the
potential profiles for extrapolation. In particular, for a finite
size system the boundary condition, $\rho(r_{\rm max})=0$, requires that no
orbit should extend beyond $r_{\rm max}$, which translates into an energy
bound, $E<-\phi(r_{\rm max})$. In other words, the energy of particles has
to be bound for a finite size system described by $f(E,L)$. 
This constraint is the main cause of the poor match between model and data for simulated DM particles in ~\citep{Wang}. This
constraint does not apply to our method, however, because we do not
need to study the full radial range of every orbit. For example, our
method can be applied to an open system with constant inflows and
outflows! In this sense, the oPDF is more general than Jeans theorem.

Fitting with a full DF also has its advantage over the general oPDF method. The prior assumptions on the distribution of orbits serve to input extra information to the model. If these assumptions are correct, the fitting can be more efficient, as demonstrated by the performance of the \emph{true} $f(E,L)$ DF fitted to the ideal tracers. On the other hand, incorrect assumptions are likely to lead to biased results in the fits. From this point of view, adding extra assumptions in the construction of a DF is a trade-off between efficiency and correctness. Our oPDF method is specifically designed to minimize extra assumptions hence maximizing correctness.

\subsubsection{Connection to Schwarzschild's method}\label{sec:Schwarzschild}

Our radial likelihood method can be regarded as a lightweight Schwarzschild's
method. Starting from the oPDF, $P(r|E,L)$, one can populate different
orbits with tracer particles given a potential, and look for weighted
combinations of orbits that reproduce the observed spatial
distribution of the tracer. The best match then gives an estimate
of both the potential and a phase-space distribution in the form of
combinations of orbits. This is the exactly the Schwarzschild
method~\citep{Schwarzschild}, which essentially converts the $\rho=\int
fd^3v$ into linear equations in phase-space grids $\rho(I)=\sum_J C(J)
P(I|J)$, where $I$ denotes configuration grids and $J$ denotes orbit
populations. However, to infer the potential, it is not necessary to
solve for a general combination of orbits, $C(J)$. Instead, one can
obtain the distribution of orbits directly from the observed phase-space
positions of particles, with each particle determining one orbit,
i.e, $C(J)=1$ with $J$ ranging from 1 to the number of particles. 
This is exactly what we do in our likelihood method. In this sense, our likelihood method is a special type
of Schwarzschild's method, with the population of orbits constrained
to be the distribution of orbits in the data, rather than constructed
from an external library. We do not lose generality with our choice of
orbits, while hugely reducing the dimension of the problem by not
solving for $C(J)$ at all.

The disadvantage of not fitting for the distribution of orbits in our method is that 
we have to rely on the full phase-space data to initialize each orbit. 
When only certain moments of the DF such as the velocity dispersion profile are available,
Schwarzschild's method can still be applied by fitting the predicted moments of 
the proposed DF to the observed ones, while the radial likelihood method constructed here cannot. 
Another advantage of the general Schwarzschild modelling is that it can be adopted to construct a self-gravitating equilibrium system, which can be used as initial conditions for N-body experiments.


\subsection{Generalization of the likelihood method}
Observational data usually involve selection functions describing
the non-uniform sample completeness, noise in the measurements of the
phase-space coordinates, and even missing dimensions in the
coordinates. We briefly discuss how these complexities can be handled
in the oPDF framework, as well as generalizations to non-spherical potentials. 
Given that the focus of this paper is to explore whether a general dynamical 
method is applicable to simulated haloes for the inference of the halo potential, 
we do not push the following discussions further to implementation. Instead, we leave
further tests and improvements of the proposed solutions to future work. In the following,
we will focus on the likelihood estimator as an example.

\subsubsection{Selection function, noise, missing dimensions}\label{sec:obs}
As we have discussed before, our methods apply to tracers with any
$(E,L)$ distribution, and hence is immune to any selection in
$(E,L)$. Radial selection can be easily handled by modifying the
reciprocal probability as
\begin{equation}\label{eq:select}
P_{ij}^\prime=\frac{P_{ij}S_i/S_j}{\int P_{ij}S_i/S_j \,dr_i},
\end{equation}
 where $S_i=S(r_i)$ is the probability of selecting a particle into the sample at
 $r_i$ . If the selection is simply a radial cut, then
 equation~\eqref{eq:select} simplifies to an adjustment of the
 normalization factor $T$. When angular selections are involved, one
 needs to explicitly consider $\vec{L}$ instead of $L$ as the orbital
 parameters, to model the distribution in $r,\theta,\phi$ rather than
 just $r$.

The noise in the data can be incorporated as priors. Formally, the new likelihood after marginalizing over the error distribution can be written as
\begin{equation}
 \mathcal{L}'_\psi(D_o)=\int \mathcal{L}_\psi(D) P(D|D_o) dD,
\end{equation} where $\mathcal{L}_\psi(D)$ is the likelihood of an error-free dataset, $D$, in a potential $\psi$, $P(D|D_o)$ is the probability that the true dataset is $D$ given the observed dataset $D_o$.
Alternatively, one
can generate Monte-Carlo realizations of the data according to the
prior distributions, $P(D|D_o)$, and
apply the method to each realization assuming no noise in the
measurements. Once this is done, a statistical estimate of the effect
of the measurement noise on the fits can be obtained from the
distribution of the best-fitting parameters across the different
realizations.

Observational data might also miss some dimensions. For example, it
is difficult to measure the tangential velocity for distant stars
in the Galaxy. If only $v_r$ is available, then it is necessary to
introduce additional assumptions on $v_t$ in order to apply the
method, e.g., through an anisotropy parameter or anisotropy profile, $\beta(r)$.

\subsubsection{Generalization to arbitrary potentials}

For a non-spherical potential, it might be difficult to write down the
integrals of motion as orbital parameters. However, the orbit is still
fully determined for each particle once a potential is assumed, and
one can calculate the orbit numerically without knowing the integrals
of motion. 
With calculable  orbits, we can still predict the
spatial distribution of particles by superimposing the oPDF of each
particle, and compare with the observed distributions for a likelihood
analysis of the potential.

\section{Conclusions}

We have shown that tracers in a steady state in a static
potential can be characterized by an orbit-dependent distribution function,
$\D P(\lambda|{\rm orbit})\propto \D t(\lambda)$, with $\lambda$ being an affine parameter of
the position along the orbit. This is a general result that follows from the
time-independent collisionless Boltzmann equation. We clarify that the
phase-space distribution of tracers connects to their host potential
only through this oPDF, while the distribution of orbits, e.g.,
$P(E,L)$, is a characteristic of each tracer that is independent of
the host potential. The oPDF can also be shown to be equivalent to
 Jeans theorem, which is the starting point for constructing DFs for
steady-state tracers in most previous studies.

Starting solely from this oPDF, we have developed a likelihood estimator
to infer the potential of a spherically symmetric halo. The method
improves over previous $f(E,L)$ DF methods in making no assumption
about the tracer characteristic functions, $P(E,L)$. We achieve this by
approximating the prior distribution of orbits, $P(E,L)$, by their
empirical distribution once a halo potential is assumed, and
marginalize over this distribution. The approximation of $P(E,L)$ by
its empirical version introduces strong shot noise, which is
suppressed by binning the data radially.

To test the performance of the likelihood estimator we have created Monte-Carlo
samples of steady-state tracers from a realistic phase-space DF for
Milky Way halo stars. The DF of these samples is constructed to be in
a steady state but also makes additional assumptions about the orbit population.
Applying our estimator to these samples, we find it to be unbiased. Comparing our
estimates with those from a likelihood estimator that uses the correct
form of the underlying full DF, our estimated errorbars are only increased
slightly ($\sim20\%$), while avoiding having to assume any functional form for
the DF. Such a likelihood estimator can be easily embedded into a Bayesian framework.

Expressed in action-angle coordinates, the oPDF reduces to the random phase principle proposed
in BL04, which has been used to construct minimum distance estimators
of the potential. When applied to the inference of an NFW
potential, the minimum distance estimators suffer from a strong
degeneracy in halo parameters, reflecting the fact that they only constrain the halo mass
inside a tracer-specific radius. While this degeneracy is an obvious disadvantage of these methods, 
it also opens a door to non-parametric reconstruction of the potential profile (or rotation curve) due to its 
independence on the shape of the proposed profile. Taking advantage of this shape-independence,
we have developed a non-parametric ``phase-mark'' method to reconstruct the potential profile, by fitting elementary
profiles to radially split subsamples of the tracer to mark the characteristic mass in each radial bin. Applied to the Monte-Carlo samples, 
we have shown that the phase-mark correctly reconstructs the true potential without making any assumptions about its shape. Such reconstructed profiles can be further fitted to provide parametric constraints on the potential. We find that the constraining power of such fits can be as good as that of the likelihood method and the tightest constraint is obtained with only two radial subsamples. The phase-mark method is more intuitive for recovering the potential profile due to its non-parametric nature. It is also computationally much faster. 

Both the likelihood estimator and the phase-mark are able to break the degeneracy between mass and concentration and constrain the shape of the halo mass profile over a large radial range. The generality of the oPDF also means that our methods can be applied to
tracers with multiple components, but without the necessity of modelling each component separately. In the current form, the new methods developed in this paper can serve as a powerful tool to study the dynamical status of simulated haloes. They also offer a promising way to constrain the mass of the Milky Way halo with real data, once further extended and tested to work with observational errors and incompleteness.

\section*{Acknowledgements}
We thank Julio Navarro, Andrew Pontzen, Vincent Eke and Yanchuan Cai for helpful comments and discussions. 
We are also grateful to the anonymous referee for enlightening comments which helped to improve the clarity of the paper.
This work was supported by the European Research Council [
GA 267291] COSMIWAY and Science and Technology Facilities Council
Durham Consolidated Grant. WW acknowledges a Durham Junior Research Fellowship.
This work used the DiRAC Data Centric system at Durham University,
operated by the Institute for Computational Cosmology on behalf of the
STFC DiRAC HPC Facility (www.dirac.ac.uk). This equipment was funded
by BIS National E-infrastructure capital grant ST/K00042X/1, STFC
capital grant ST/H008519/1, and STFC DiRAC Operations grant
ST/K003267/1 and Durham University. DiRAC is part of the National
E-Infrastructure. This work was supported by the Science and
Technology Facilities Council [grant number ST/F001166/1]. 

The code implementing our method is freely available on GitHub at \url{http://kambrian.github.io/oPDF/}.

\bibliographystyle{\mybibstyle}
\setlength{\bibhang}{2.0em}
\setlength\labelwidth{0.0em}
\bibliography{fitting}

\appendix
\section{AD distribution}\label{app:ADdist}
\begin{figure}
 \myplot{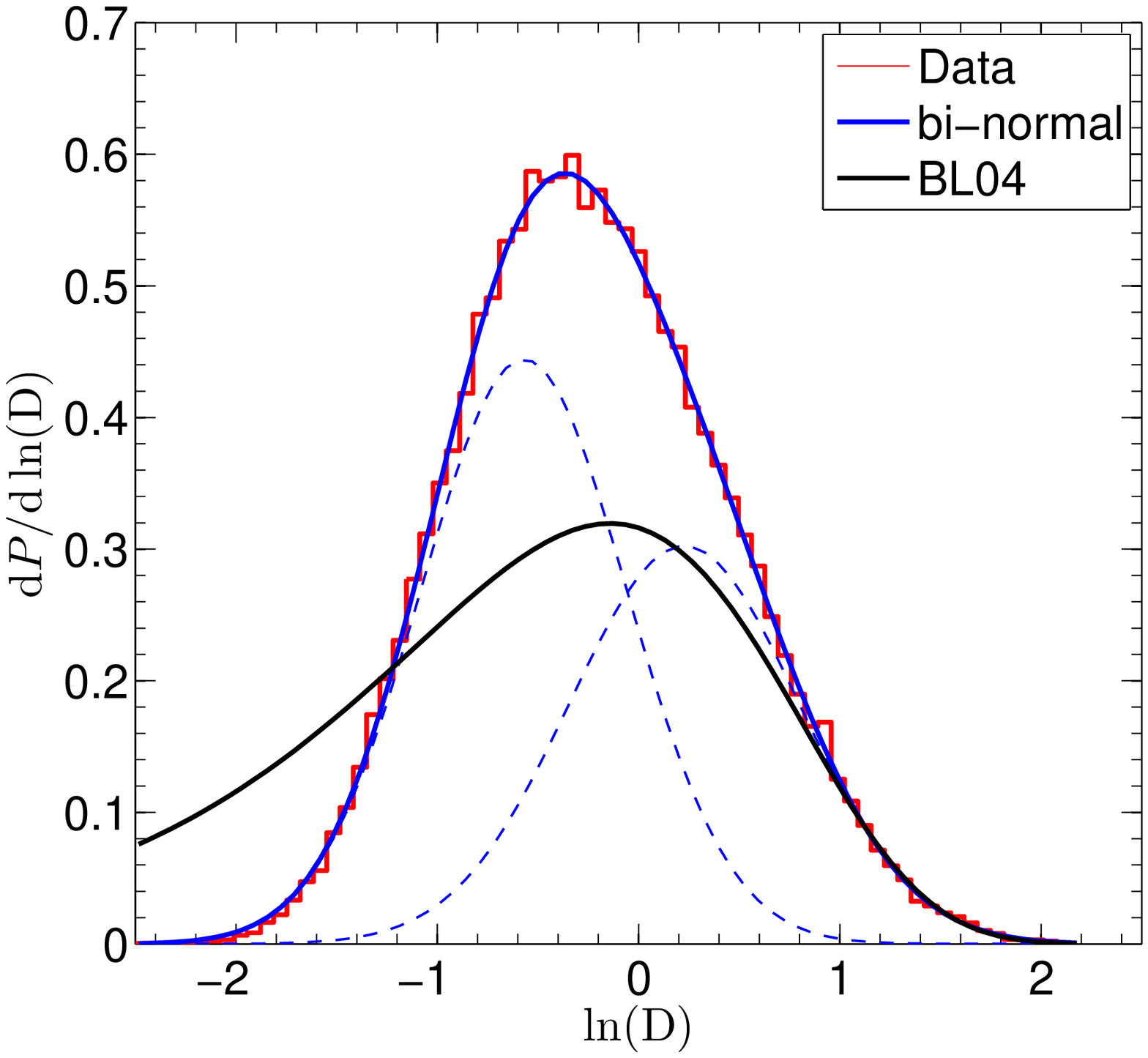}
 \caption{Binormal fit to the distribution of the AD statistic. The
 data are generated from an ensemble of $50\,000$ random samples, of
 size $N=5000$ each. For each of these samples, the AD statistic, $D$, is
 calculated. The empirical distribution of $\ln(D)$ is
 plotted as a histogram. The solid line passing through the histogram is a bi-normal fit according to Equation~\eqref{eq:ADfit}, which is the sum of two Gaussian components (dashed lines). For comparison, we also plot the BL04 fit which is only
 designed to describe the tail of the distribution.}\label{fig:AD}
\end{figure}

The theoretical distribution of the AD distance (Eq.~\ref{eq:AD}) under the null hypothesis can be calculated with Monte-Carlo simulations. Specifically, we generate a number of independent random samples, and calculate the AD distance, $D$, for each of them. Each sample consists of $N$ independent observations of a uniformly distributed variable $\theta \in [0,1]$. In Fig.~\ref{fig:AD}, we show that the distribution of $\ln(D)$ (Eq.~\ref{eq:AD}) can be well fit by the sum of two normal distributions of the form
\begin{equation}\label{eq:ADfit}
 P\left(\ln{D}\right)=w \mathcal{N}\left(\ln{D}, \mu_1, \sigma_1\right)+(1-w) \mathcal{N}\left(\ln{D}, \mu_2, \sigma_2\right),
\end{equation}
where $\mathcal{N}\left(x,\mu,\sigma\right)$ is the standard normal probability
function of $x$ with mean $\mu$ and standard deviation $\sigma$. The best fit
parameters are $w=0.569$, $\mu_1=-0.570$, $\sigma_1=0.511$,
$\mu_2=0.227$, $\sigma_2=0.569$. Compared with the fitting function in
BL04 designed to fit the tail of the distribution, the bi-normal PDF 
fits well the whole range of the distribution, which is important for
likelihood analysis.

The distribution has barely any dependence on the sample size $N$. For systems as small as $N=5$, we find our fitting still describes the empirical AD distribution very well. We also verified that the mean phase distribution can be well approximated by the normal distribution, for systems with $N\geq 5$.

%

\end{document}